\documentclass{emulateapj}

\def\mj{M$_{\rm J}\ $}
\def\rj{R$_{\rm J}\ $}
\def\etal{{et~al.\,}}

\def\teff{T$_{\rm eff}\,$}

\def\mic{$\mu$m$\,$}

\def\sles{\lower2pt\hbox{$\buildrel {\scriptstyle <}
   \over {\scriptstyle\sim}$}}
\def\sgreat{\lower2pt\hbox{$\buildrel {\scriptstyle >}
   \over {\scriptstyle\sim}$}}

\begin{document}

\slugcomment{Submitted to Ap.J. April 29, 2006; Accepted June 26, 2006}

\title{Theory for the Secondary Eclipse Fluxes, Spectra, Atmospheres, 
and Light Curves of Transiting Extrasolar Giant Planets} 

\author{A. Burrows\altaffilmark{1}, D. Sudarsky\altaffilmark{1} \& I. Hubeny\altaffilmark{1}} 

\altaffiltext{1}{Department of Astronomy and Steward Observatory, 
                 The University of Arizona, Tucson, AZ \ 85721;
                 burrows@zenith.as.arizona.edu, sudarsky@as.arizona.edu, hubeny@aegis.as.arizona.edu}

\begin{abstract}

We have created a general methodology for calculating the 
wavelength-dependent light curves of close-in extrasolar 
giant planets (EGPs) as they traverse their orbits. Focussing 
on the transiting EGPs HD189733b, TrES-1, and HD209458b, we 
calculate planet/star flux ratios during secondary eclipse 
and compare them with the {\it Spitzer} data points obtained 
so far in the mid-infrared. We introduce a simple 
parametrization for the redistribution of heat to the 
planet's nightside, derive constraints on this parameter 
(P$_n$), and provide a general set of predictions for 
planet/star contrast ratios as a function of wavelength, 
model, and phase. Moreover, we calculate average dayside 
and nightside atmospheric temperature/pressure profiles 
for each transiting planet/P$_n$ pair with which existing 
and anticipated {\it Spitzer} data can be used to probe the 
atmospheric thermal structure of severely irradiated EGPs.
We find that the baseline models do a good job of fitting 
the current secondary eclipse dataset, but that the 
{\it Spitzer} error bars are not yet small enough to 
discriminate cleanly between all the various possibilities.

\end{abstract}

\keywords{stars: individual (TrES-1, HD209458, HD189733)---(stars:) planetary systems---planets and satellites: general}

\section{Introduction}
\label{intro}

Probing the atmospheres of extrasolar giant planets (EGPs) by measuring
their spectra is the paramount means to determine their physical and chemical
character (Burrows 2005).  Such direct measurements complement the kinematic and
orbital information obtained through the radial-velocity (RV) technique by
which the vast majority of the EGPs have to date been discovered and 
studied. However, an EGP's spectrum and phase-dependent light curve can 
in principle reveal or constrain the molecular and atomic
compositions, atmospheric temperatures, cloud properties, 
albedos in the optical, and the degree to which the
heat absorbed on the dayside is redistributed to the 
nightside before reradiation.  The advection of heat and material 
from the dayside by jet streams and zonal winds will alter the dayside
atmospheric temperatures and non-equilibrium compositions (Menou et al. 2002;
Cho et al. 2003; Burkert et al. 2005; Iro, B\'ezard, \& Guillot 2005; Showman \& Guillot
2002; Guillot \& Showman 2002), and might measurably shift
the light curve with respect to the orbital ephemeris 
(Cooper \& Showman 2005; Williams et al. 2006).  Dynamic meteorology could also introduce
zonal banding, as seen in Jupiter and Saturn, and temporal fluctuations, 
and does influence the rate with which heat is lost from the inner core 
(Burrows, Sudarsky, \& Hubbard 2003; Burrows et al. 2004), 
and thereby the radius of the planet and its evolution. 
Moreover, such redistribution affects the near- and mid-infrared emissions 
from the night side, and as a result will affect the interpretation of nightside data 
when they become available.

The planet/star flux ratios of wide-separation EGPs ($>0.2$ AU) 
are quite low ($10^{-4}$ to 10$^{-14}$) and vary widely
as a function of wavelength and orbital separation 
(with the concomitant non-monotonic variations in 
geometric and Bond albedos) (Sudarsky, Burrows, \& Pinto 2000; Burrows, 
Sudarsky, \& Hubeny 2004; Sudarsky et al. 2005; Burrows 2005).  Nevertheless,
space-based coronagraphic techniques can be designed with inner working 
angles and contrast capabilities that will eventually
image such planetary systems in the optical and mid-IR and 
distinguish planet from star (Trauger et al. 2000; 
Trauger, Hull, \& Redding 2001).   However, the close-in 
EGPs with orbital semi-major axes less than $\sim$0.1 AU will 
not be imaged separately any time soon.  In these cases, 
distinguishing the planet's spectrum from that of the star
requires different techniques that don't rely on imaging.

Fortunately, it has been shown recently by Charbonneau
et al. (2005) and Deming et al. (2005,2006) that the {\it Spitzer} 
infrared space telescope can discern changes in the summed
light of a transiting EGP and primary star due to the occultation
of the planet by the star during secondary eclipse (phase 
angle, $\alpha$, near 0$^{\circ}$).  The difference 
in the summed light just before and during 
planetary eclipse provides a measure of the irradiated 
planet's emissions in the {\it Spitzer} IRAC bands at 3.6 \mic, 4.5 \mic,
5.8 \mic, and 8.0 \mic, in the MIPS band at 24 \mic, and via the {\it Spitzer}/IRS.
To date, nine transiting EGPs have been discovered 
(Charbonneau, Brown, Burrows, \& Laughlin 2006), four
(HD209458b, TrES-1, HD189733b, and HD149026b) are close enough
to attempt secondary eclipse measurements with adequate signals-to-noise, 
and, as of this writing, eclipses for three transiting EGPs have in 
fact been detected\footnote{The planet/star flux contrast ratio for 
HD149026b may, however, be too low for a successful {\it Sptizer}
campaign.}. The corresponding planet/star flux ratios\footnote{actually, detected electron ratios} 
at superior conjunction are for TrES-1 0.00066$\pm$0.00013 and 0.00225$\pm$0.00036 at 4.5 \mic 
and 8.0 \mic, respectively (Charbonneau et al. 2005), for HD209458b 
0.0026$\pm$0.00045 at 24 \mic (Deming et al. 2005), and for HD189733b 
0.0055 $\pm$0.00017 at $\sim$16 \mic in the IRS peak-up band (Deming et al. 2006). 
Along with the inferences using HST/STIS of the presence of sodium 
in the atmosphere of HD209458b (Charbonneau et al. 2002; 
Fortney et al. 2003; Allard et al. 2003) and of photolytic atomic 
hydrogen in its wind (Vidal-Madjar et al. 2003), these data are the
first direct ``spectral" measurements of extrasolar planet atmospheres
and are early harbingers of the numerous programs of EGP
remote sensing from the ground and from space being planned and/or
proposed.

The secondary eclipse data for HD209458b and TrES-1 have been subjected
to preliminary theoretical analysis by four groups.  Burrows, Hubeny, \& Sudarsky (2005)
concluded that these data are best interpreted with atmospheres
containing water and carbon monoxide for which redistribution to the night side
is significant, but partial.  They conclude that the metallicity
dependence is very weak and predict that the flux at 3.6 \mic is 
higher than that at 4.5 \mic.  They also predict a broad peak near 10 \mic,
not so obvious in the theoretical results of others. Seager et al. (2005) emphasize the potential
effects of non-solar C/O ratios above 1.0, in particular the associated
lowering of the water abundance and weakening of the water absorption features.
They also suggest that the dayside reradiates most of the stellar heat absorbed
and incorporate into their arguments the upper limit near 
2.2 \mic found for HD209458b by Richardson, Deming, \& Seager (2003).
Fortney et al. (2005) have trouble fitting the steep spectral slope 
seen in TrES-1 between 4.5 \mic and 8.0 \mic, without a significant enhancement
in metallicity.  With enhancements of 3 to 5, they fit the two TrES-1 
data points to within 2-$\sigma$ (8 \mic) and 1-$\sigma$ (4.5 \mic).
Furthermore, for both TrES-1 and HD209458b they prefer uniform 
reradiation of the absorbed stellar light over the entire planetary 
sphere and, hence, complete heat redistribution.  Barman et al. (2005)
calculate 2D planetary atmospheres and a set of light curves 
for TrES-1 and HD209458b and redistribute heat from the dayside with a redistribution 
factor $f$ (Burrows et al. 2000; Burrows, Sudarsky, \& Hubbard 2003),
also used by Burrows, Hubeny, \& Sudarsky (2005) and Fortney et al. (2005). They
assume nightside core fluxes consistent with fixed values of effective temperature (\teff) 
of 225 K and 500 K for TrES-1 and HD209458b, respectively.
Barman et al. (2005) conclude that some redistribution must be occurring in the 
atmospheres of both HD209458b and TrES-1, but have trouble simultaneously fitting the 
two TrES-1 data points.  All groups find that the atmospheric temperatures
are, as expected, hot and above $\sim$1000 K, but the predicted 
planet/star contrast ratio spectra at superior conjunction vary 
perceptibly from group to group.  

With the IRS peak-up measurement of HD189733b at $\sim$16 \mic, 
and the secondary eclipse data anticipated in the near future 
for the remaining combinations of object and {\it Spitzer} band
\footnote{bringing the total number of data points or constraints 
around secondary eclipse to 18 (!)}, the time is ripe for a new set of 
theoretical spectral models and predictions for HD209458b,
TrES-1, and HD189733b at superior conjunction ($\alpha = 0$), as well as for 
the corresponding light curves for the general phase angle, $\alpha$. 
In this paper, we provide such models and compare to the extant data
to extract physical information about the atmospheres of these three
transiting EGPs.  We also make predictions for the light curves as a function
of wavelength and the degree of redistribution to the nightside, and 
explore the metallicity dependence of the secondary eclipse predictions.
To calculate the phase light curves, we use the 2D photon transport code
and technique described in Sudarsky et al. (2005), but introduce the 
redistribution parameter, P$_{n}$, which is the fraction of the stellar
energy intercepted by the planet that is redistributed to the nightside  
\footnote{Note that with this definition, if the Bond albedo
were large (which is the case only for cloudy models we don't discuss in this paper),
the P$_n$ = 0.5 model would result in slightly greater IR fluxes from the nightside
than the dayside.}. P$_{n}$ = 0 means no redistribution. We calculate for 
a given star/planet system and P$_{n}$ both the dayside and nightside atmospheric
temperature($T$)/pressure($P$) profiles and the associated spectra, and then
for a given phase angle, $\alpha$, combine the emissions from the two
hemispheres to derive the total planet fluxes at the Earth 
for 300 wavelengths logarithmically spaced from 2.5 
\mic to 30 \mic and the corresponding planet/star flux 
ratios.  Limb darkening effects for the day and night sides 
and planetary Bond, geometric, and spherical albedos for the 
day side are automatically derived in the calculations and are 
not imposed artificially. 

We have opted in this paper for the P$_n$
parametrization, and not the $f$ parametrization mentioned above
and introduced by Burrows et al. (2000), because it is better
tied to the core issue of heat redistribution, and because $f$ is definitionally
tied to stellar irradiation, which is in fact absent on the nightside. Modeling
the heating on the nightside with a flux at the base of the atmosphere that
accounts for the advection of heat by winds seemed a bit more physical
than heating the nightside by the ensatz of external insolation.
The results are predictions 
for the three transiting EGPs as a function of wavelength or 
{\it Spitzer} band and six values of P$_{n}$ (\{0,0.1,0.2,0.3,0.4,0.5\}).
We thereby derive the dependence of the planet/star flux ratio spectra 
upon redistribution fraction and phase angle, albeit in the context 
of a simplified meteorological model. The spectral model, however, 
is state-of-the-art.  In addition, we determine for the 
three close-in EGPs the approximate P$_{n}$ dependence of the day 
and night side $T/P$ profiles and pay special attention to the 
temporal and phase dependence of the flux ratios in the IRAC bands and the 24-\mic 
MIPS band during a full orbit ($0^{\circ} < \alpha < 180^{\circ}$),
not just at superior conjunction (secondary eclipse).
In this way, we provide a complete set of theoretical models both for 
comparison with current data and for predicting future measurements. 
Though our baseline models are for solar metallicity, we find that the metallicity
dependence, without clouds and for solar abundance ratios, is small (see \S\ref{comparison}).

\section{Why the Mid-Infrared is Best and New Theoretical $T/P$ Profiles}
\label{why}

\begin{figure}
\epsscale{1.00}
%\vspace*{-0.7in}
\plotone{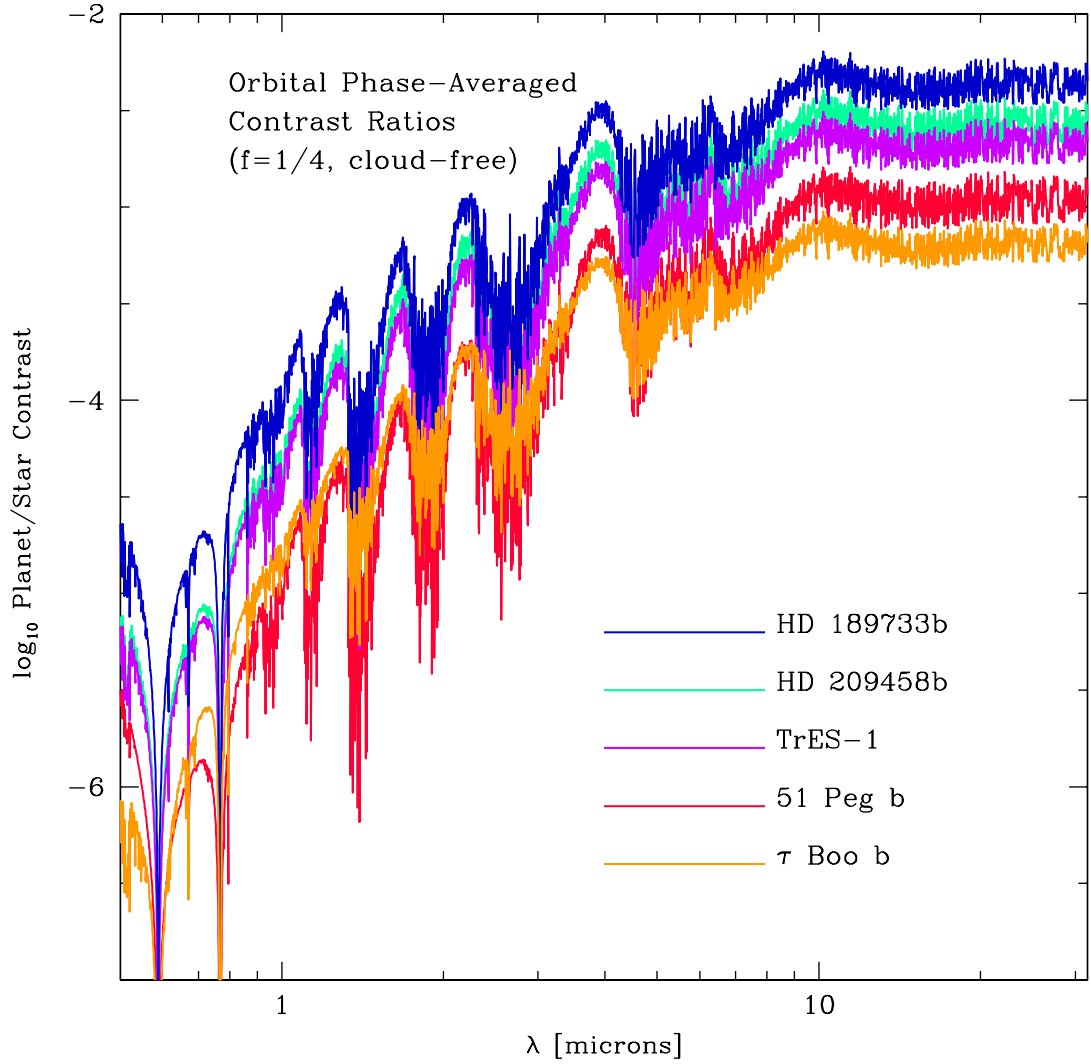}
%\vspace*{-0.2in}
\caption{Logarithm of the planet/star flux density contrast ratio versus wavelength
from 0.5 \mic to 30 \mic, for five close-in EGPs: HD189733b, HD209458b,
TrES-1, 51 Peg b, and $\tau$ Boo b.  These ratios have been averaged over
a full orbit, assume complete redistribution of heat using the old prescription
($f = 0.25$; Burrows, Sudarsky, \& Hubbard 2003), and are from cloud-free models.
Note the variation of more than three orders-of-magnitude from the optical to
the mid-IR, where, all else being equal, the ratios are most favorable and {\it Spitzer}
is currently the platform of choice. See text for a discussion.
\label{fig:1}}
\end{figure}

It is in the near- and mid-infrared that the planet-to-star flux ratio
is most favorable for the direct detection of the light of a close-in 
EGP.  The general theory makes this clear (Burrows et al. 2001; Burrows et al. 2004;
Sudarsky et al. 2005; Burrows 2005), but this can be most easily demonstrated
with a graph of the contrast ratio versus wavelength for a sample of
the closest EGPs.  Figure \ref{fig:1} depicts theoretical 
orbital-phase averaged (Sudarsky, Burrows, \& Hubeny 2003)
flux ratios from the optical through 30 \mic, under 
the assumption of complete redistribution and ignoring
any possible cloud effects, for five of the most interesting
close-in EGPs: HD209458b, HD189733b, TrES-1, 51 Peg b, and $\tau$ Boo b. 
The first three are transiting and are discussed in more detail in following sections.

Figure \ref{fig:1} demonstrates the advantages of working in the mid-IR.
Not only is the stellar flux lower there, but absorption by nascent sodium and potassium
in the optical and near-IR renders measurements at shorter wavelengths problematic,
even for the most sensitive space telescopes.  
To date, only upper limits have been obtained to the 
planet/star contrasts in the optical and the corresponding
albedos (e.g., Rowe et al. 2006). Contrasts in
the optical of $10^{-5}$ to $10^{-6}$ are expected, 
improving to only $10^{-4}$ to 10$^{-3}$ in the near-IR.

However, as Fig. \ref{fig:1} suggests, in the mid-IR the planet/star contrasts rise
above $10^{-3}$, making {\it Spitzer}, and later JWST, the preferred
platforms for the spectral study of close-in and transiting giant planets. 
Note that there is a full order-of-magnitude range in the predicted
average contrast ratio in the mid-IR, reflecting predominantly the range of 
stellar fluxes and orbital separations. Fortuitously, 
the transiting EGPs upon which we focus in this paper have the most
favorable ratios of the set.

Table \ref{EGP_tab} lists the physical parameters we assume for the
three transiting EGPs and their primaries.  The most important
are the orbital radius ($a$), period (P), planetary radius (R$_p$),
and planetary mass (M$_p$), along with the stellar properties that
determine the stellar luminosity and spectrum.  In addition, Table \ref{EGP_tab} 
gives the surface gravity ($g_p$) employed for each EGP.  The 
stellar irradiation spectra are taken from Kurucz (1994) 
and we use the 1D and 2D atmosphere and irradiation codes outlined 
in Burrows, Sudarsky, \& Hubeny (2004) and Sudarsky, Burrows, \& 
Hubeny (2005).

\begin{figure}
\epsscale{1.00}
%\vspace*{-0.7in}
\plotone{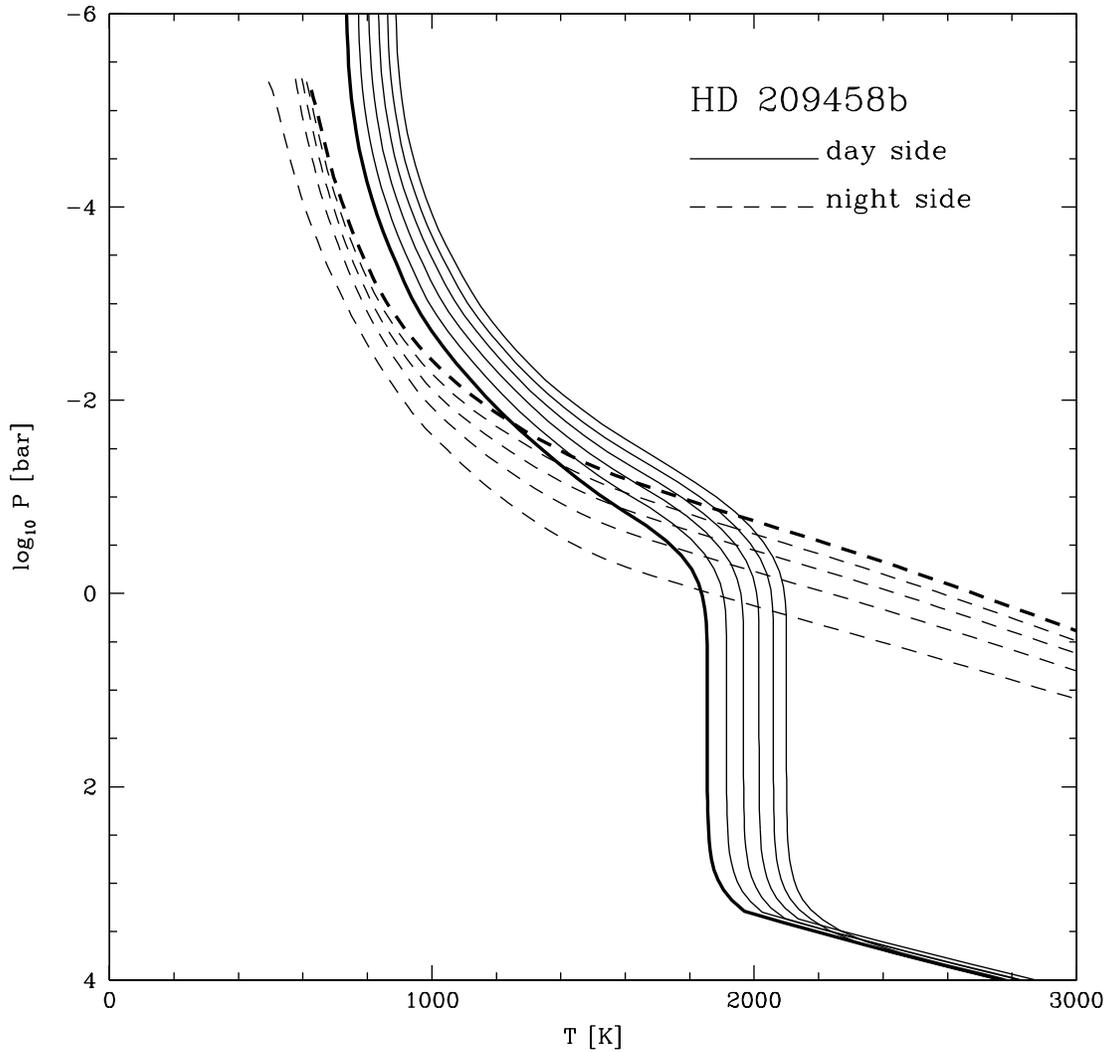}
%\vspace*{-0.2in}
\caption{Theoretical profiles of the logarithm base ten of the pressure 
($P$, in bars) versus temperature ($T$, in Kelvin) for both the dayside 
(solid) and the nightside (dashed) atmospheric models of HD209458b, for P$_n$s from 0.0 to
0.5 in steps of 0.1.  The P$_n$ = 0 model for the nightside is omitted.  The thicker lines
are for the P$_n$ = 0.5 models.  See text for details and discussion.
\label{fig:2}}
\end{figure}

\begin{figure}
\epsscale{1.00}
%\vspace*{-0.7in}
\plotone{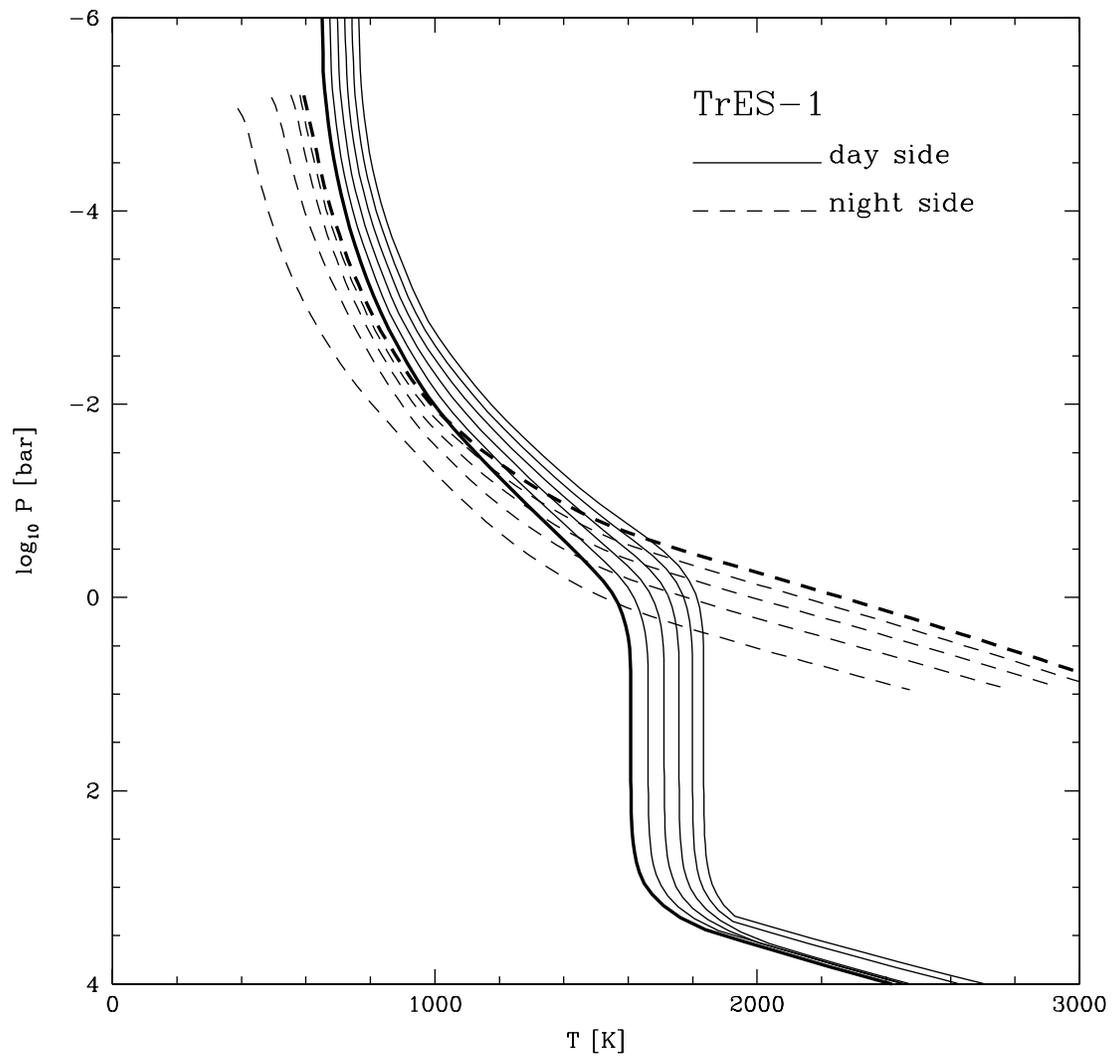}
%\vspace*{-0.2in}
\caption{Same as Fig. \ref{fig:2}, but for TrES-1.
\label{fig:3}}
\end{figure}

\begin{figure}
\epsscale{1.00}
%\vspace*{-0.7in}
\plotone{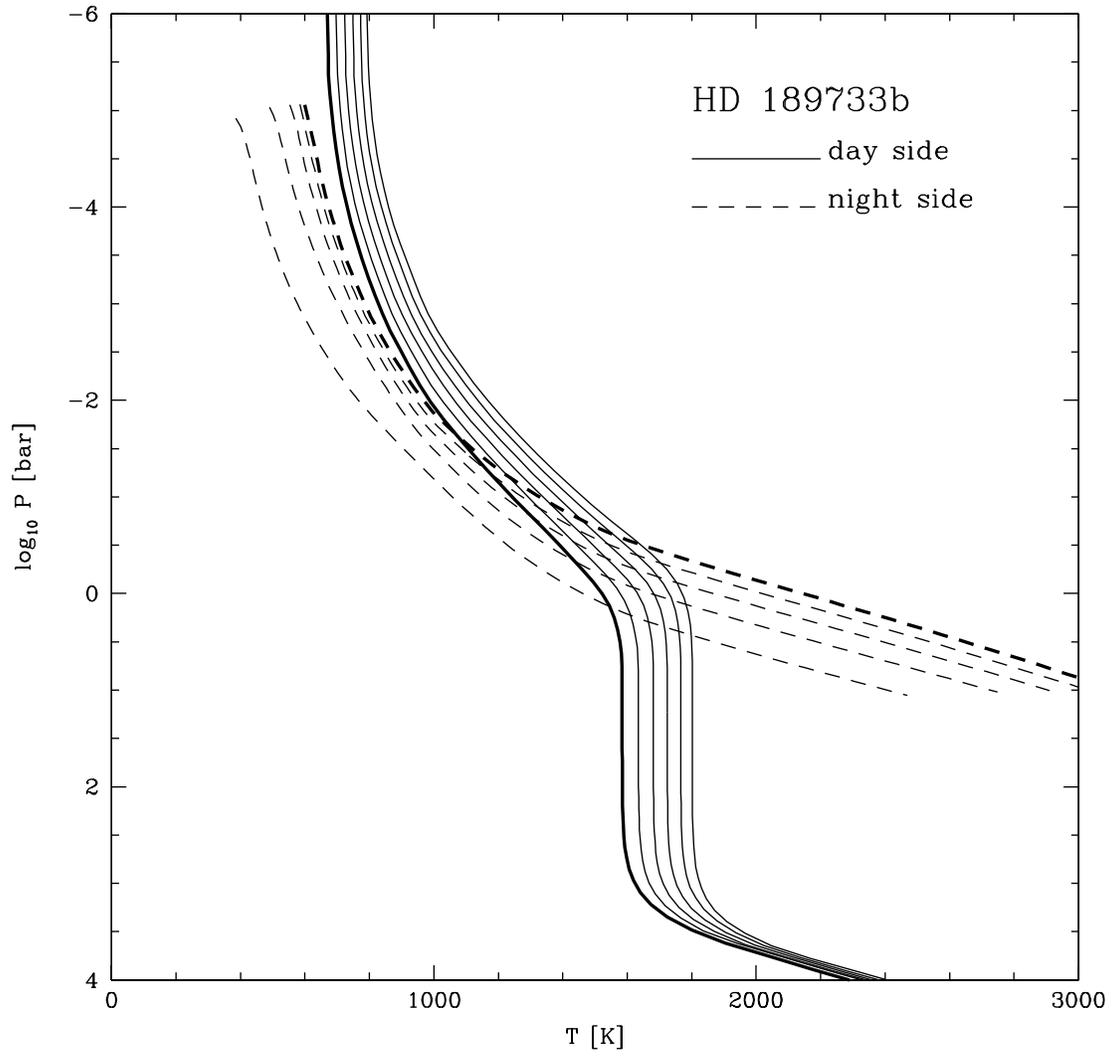}
%\vspace*{-0.2in}
\caption{Same as Figs. \ref{fig:2} and \ref{fig:3}, but for HD189733b.
\label{fig:4}}
\end{figure}

Figures \ref{fig:2}, \ref{fig:3}, and \ref{fig:4} depict the $T/P$
profiles for both the day (solid) and night (dashed) sides of our models of HD209458b, 
TrES-1, and HD189733b for values of P$_n$ from 0.0 (or 0.1) to 0.5, in steps of 0.1.
P$_n$ = 0 implies no redistribution to the nightside and P$_n$ = 0.5
assumes that 50\% of the stellar energy intercepted by the planet is advected
to the night side, where it is radiated. The thicker line of 
the two hemisphere sequences for each EGP is for P$_n$ = 0.5; as 
Figs. \ref{fig:2}, \ref{fig:3}, and \ref{fig:4} indicate, the dayside
atmospheres are hotter for lower P$_n$s, while the nightside atmospheres
are cooler for lower P$_n$s, as would be expected.  Since for P$_n$ = 0 the
nightside atmosphere is so much cooler than for the other values, P$_n$ = 0 
nightside $T/P$ profiles are not shown. 

Without clouds\footnote{In constructing the models present in this paper we have neglected clouds.
This is sensible for TrES-1 and HD189733b, due to the fact that their
$T/P$ profiles do not intercept at altitude the silicate or iron condensation curves.
However, for HD209458b, clouds in its upper atmosphere may well have some effect. We plan to revisit
this issue in a following paper.}, the albedos on the dayside
are due solely to Rayleigh scattering.  As a consequence, the Bond albedos
are low ($<0.1$; Sudarsky, Burrows, \& Pinto 2000) and P$_n$ = 0.5
corresponds approximately to symmetric losses on the day and night sides.
We point out that the roughly isothermal region on the dayside
is an effect of stellar irradiation, which is absent on the nightside.
This is the region where the outward core flux and the inwardly penetrating stellar 
flux are in rough balance (see Hubeny, Burrows, \& Sudarsky 2003 for 
a more detailed explanation). It is important to note that despite the fact that the 
inner boundary condition for EGP atmospheres on both the day and the night sides
should be that they have the same interior entropy and surface gravity, 
the nightside and dayside atmospheres depicted in Figs. \ref{fig:2}, \ref{fig:3}, 
and \ref{fig:4} clearly have different interior entropies.  This is because the
advection of heat to the nightside by winds and jet streams is an additional heat
source.  For the purposes of this study, we place this nightside source at the 
base of the nightside atmosphere, which as the Figures show is at
much lower pressures than the convective zone on the dayside.  
Given that we have no model for the 3D general circulation, this is a reasonable ansatz
and gives the proper emergent fluxes.  The associated nightside \teff
is given by the formula:

\begin{equation}
{\rm T}_{\rm eff}^4 = {\rm P}_n \Omega {\rm T}_{\rm eff}^4({\rm star}) + {\rm T}_{\rm eff}^4({\rm intrinsic})\, ,
\label{eq:1}
\end{equation}
where $\Omega$ is the stellar flux dilution factor at the planet surface, ${\rm T}_{\rm eff}$(star)
is the star's effective temperature, and $\sigma {\rm T}_{\rm eff}^4$(intrinsic) is the planet's
core flux.  $\sigma$ is the Stefan-Boltzmann constant and ${\rm T}_{\rm eff}({\rm intrinsic})$ 
is assumed to be 50 K for the three EGPs.  At this value of ${\rm T}_{\rm eff}({\rm intrinsic})$, 
and for reasonable values (50-75 K), its effects on the emergent planet fluxes and the planet/star 
flux ratios are negligible, except on the nightside for P$_n$ = 0.0.

As Figs. \ref{fig:2}, \ref{fig:3}, and \ref{fig:4} indicate, the atmosphere of
HD209458b is hotter than those of TrES-1 and HD189733b.  Furthermore, the
atmospheric profiles of TrES-1 and HD189733b are quite similar.  On HD209458b's 
dayside and at a $P$ of $\sim$1 bar, $T$ ranges between 1800 K and 2100 K,
while the corresponding $T$s of TrES-1 and HD189733b range between 1500 K and 1900 K.  
At pressures of $\sim$0.1 bar, the difference in $T$ between the night and day 
sides of HD209458b ranges from +800 K to -200 K for P$_n$ = 0.1 and P$_n$ = 0.5, 
respectively, representing a wide, but reasonable, range (Showman \& Guillot 2002; 
Guillot \& Showman 2002; Iro, B\'ezard, \& Guillot 2005).  
The corresponding numbers for TrES-1 and HD189733b
are a bit smaller, +500 K to {$-200$} K.  For all three transiting EGPs, one
temperature does not represent the atmosphere very well.  In particular, 
the temperatures at the wavelength-dependent photospheres range by more 
than 500 K. Hence, one should be cautious when inverting a ``brightness" 
temperature to obtain an ``atmospheric" temperature.

\section{Orbital Variation of Planet/Star Flux Ratios in the {\it Spitzer} Bands}
\label{bands}

Given these atmospheric $T/P$ profiles, we turn next to the corresponding
planet/star flux ratios in the {\it Sptizer} bands as a function of orbital phase.
Figures \ref{fig:5}, \ref{fig:6}, \ref{fig:7}, \ref{fig:8}, and \ref{fig:9}
portray the light curves in the IRAC-1 (3.6 \mic), IRAC-2 (4.5 \mic), IRAC-3 (5.8 \mic),
IRAC-4 (8.0 \mic), and MIPS (24 \mic) bands for HD189733b, HD209458b, and TrES-1.
For simplicity, we have assumed orbital inclinations of 90$^{\circ}$ for all three EGPs
and we do not excise the tens of minutes during which the planet 
is actually occulted and the ratio would drop to zero. To construct our models, 
we have used the latest EGP radii for HD189733b (Bakos et al. 2006) and HD209458b 
(Knutson et al. 2006), both of which have trended down since discovery. For each 
EGP, the secondary eclipse begins near $\alpha \sim 7^{\circ}$, which due to 
the flatness of the curves near $\alpha = 0$ has the same planet/star ratios as
at $\alpha = 0$ (superior conjunction).  Figures \ref{fig:5} through \ref{fig:9}
provide the entire orbital evolution of the contrast ratios of all three transiting planets,
not just the values during secondary eclipse.  The plots are versus orbit phase,
or fraction of an orbital period, and the individual periods are given in Table 
\ref{EGP_tab} and on the figures.  As expected, the highest contrast ratios at
secondary eclipse are for the P$_n$ = 0.0 models, which also manifest the
largest temporal variation.  In addition, for a given P$_n$, the highest
values are for HD189733b, and in the four IRAC bands those for TrES-1 and 
HD209458b are similar. For the MIPS band, the TrES-1 values
exceed those for HD209458b by $\sim$20-30\%.  For P$_n$ = 0.0, the variation
with phase about the orbit mean for all five bands is about a factor of two, while that
for P$_n$ = 0.5 is $\sim$5-10\%.  There is temporal variation even for the P$_n$ =0.5 models
because the $T/P$ profiles on the day and night sides are different, despite
comparable total IR reradiation, and because the dayside optical albedos are 
non-zero.

As Figs. \ref{fig:5} through \ref{fig:9} indicate, the planet/star flux ratios
in the IRAC bands vary with P$_n$ by about 30-60\%, depending upon object and band,
but only by about 20-30\% in the 24-\mic MIPS band.  The MIPS band shows less variation
with P$_n$ at secondary eclipse because a change of P$_n$ from 0.0 to 0.5 represents 
a factor of $\sim$2 decrease in the heating of the dayside and this translates into a
$\sim 2^{1/4} - 1$ $\sim$20\% variation in the ``emission" temperature.  Since 24 microns 
is on the Rayleigh-Jeans tail, this is also the corresponding flux variation.

Figure \ref{fig:9} shows that the predicted flux ratio in the MIPS band at secondary eclipse 
varies for HD189733b between $\sim$0.55\% and $\sim$0.7\%, while those for TreS-1
and HD209458b vary from  $\sim$0.4\% to $\sim$0.47\% and $\sim$0.3\% to $\sim$0.4\%,
respectively.  These large values, very different from the low values
anticipated in the optical (Fig. \ref{fig:1}), have motivated (see also 
Burrows, Sudarsky, \& Hubeny 2003) the ongoing {\it Spitzer} observing campaigns.

\section{Planet/Star Flux Ratio Spectra as a Function of Phase and Redistribution}
\label{spectra}

Expanding from the narrower focus on the IRAC and MIPS bands in \S\ref{bands}, 
Figs. \ref{fig:10}, \ref{fig:11}, and \ref{fig:12} portray the planet/star 
contrast spectra from 3.0 \mic to $\sim$27 \mic for a representative subset 
of P$_n$ values and for HD209458b, TrES-1, and HD189733b, respectively. Shown
are these spectra for four different phase angles (0$^{\circ}$, 60$^{\circ}$, 
120$^{\circ}$, and 180$^{\circ}$) and for the three P$_n$s.  These results are consonant 
with the band light curves given in Figs. \ref{fig:5} to \ref{fig:9}, but render the
full spectral dependence on P$_n$, transiting EGP, and orbital phase.  Generically,
we find peaks near $\sim$4 \mic and $\sim$10 \mic and steeper slopes from 5 \mic to 
10 \mic near superior conjunction. For P$_n$ = 0.5, the 4-\mic peak to 10-\mic peak 
flux ratio increases with increasing phase angle away from $\alpha = 0$, but for 
P$_n = 0$ it generally decreases.  From 14 \mic to 30 \mic, the contrast spectra
are rather flat, implying that the 16-\mic IRS peak-up and the 24-\mic MIPS numbers
should be comparable.  At 60$^{\circ}$ and 120$^{\circ}$, the models with different
P$_n$s, while different, are most similar.  Naturally, the spectra at 180$^{\circ}$ are
entirely due to the nightside.  At a given wavelength, the magnitudes of the 
phase and temporal variations are similar to those discussed in \S\ref{bands} for 
those IRAC or MIPS bands that are closest in wavelength.  Figures \ref{fig:5} 
through \ref{fig:12} summarize our theoretical calculations and predictions.

\section{Comparison of Theory with Data at Secondary Eclipse and Conclusions}
\label{comparison}

\begin{figure}
\epsscale{1.00}
%\vspace*{-0.7in}
\plotone{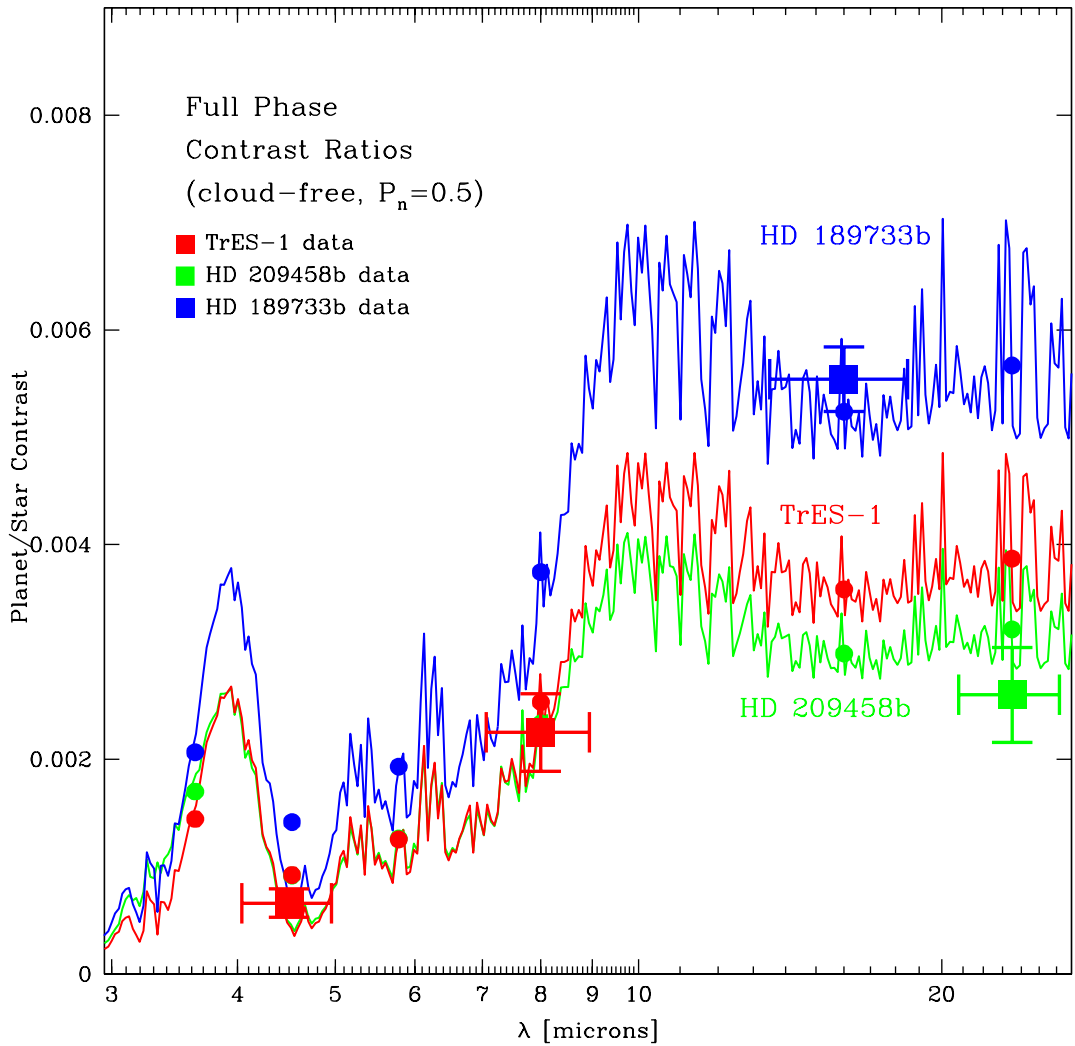}
%\vspace*{-0.2in}
\caption{Theoretical planet/star flux ratio spectra during 
secondary eclipse for wavelengths from 3 \mic to 27 \mic, 
for P$_n$ = 0.5 ($\sim$complete heat redistribution) and 
for the three EGPs HD189733b (blue), TrES-1 (red), and HD209458b (green).
Values for 300 wavelength points spaced logarithmically are included in each curve.
Superposed as large squares with 1-$\sigma$ error bars and in the 
appropriate color are the current measurements at secondary eclipse (Deming et al. 
2005, 2006; Charbonneau et al. 2005). The smaller filled circles
are the predictions for the four IRAC bands, the MIPS band at 24-\mic, and the
IRS peak-up band near 16 \mic.  To derive these numbers
we have performed a band-average of the ratio of the detected electrons.
One should compare the small dots with the large squares of the same 
color and in the same wavelength band to draw conclusions about the 
fidelity of the models and the character of the associated EGP atmospheres.
See the text in \S\ref{comparison} for a discussion of some of the inferences drawn 
from the current dataset.
\label{fig:13}}
\end{figure}

We now turn to specific comparisons between the extant {\it Spitzer} data
and our theoretical results.  Figure \ref{fig:13} portrays the planet/star
flux ratios versus wavelength at superior conjunction for 
P$_n$ = 0.5 ($\sim$complete redistribution) and the three transiting 
EGPs: HD189733b (blue), TrES-1 (red), and HD209458b (green).   
Superposed as large squares with 1-$\sigma$ flux error bars are the four 
secondary eclipse measurements to date (two for TrES-1, one for HD209458b, 
and one for HD189733b).  Also included as round dots in the appropriate color are 
the band-integrated detected electron ratio predictions for these EGP models, 
with approximate band widths indicated and no error bars in the flux direction. 

A comparison between the measured points and corresponding theoretical points
for this P$_n$ = 0.5 model is encouraging, particularly for the TrES-1 data 
at 4.5 \mic and 8.0 \mic, but also for the HD189733b IRS peak-up data point  
near 16 \mic. The 24-\mic point for HD209458b is within about 1-$\sigma$,
but slightly below the theory.  Since these P$_n$=0.5 models yield the lowest
theoretical values for the contrast ratios among the set from 0.0 to 0.5, for the HD209458b 
24-\mic point this may be the best we can do currently.  For the HD189733b point at 16 \mic, 
the entire P$_n$ range studied would still be consistent to within the 1-$\sigma$
range quoted (see the top left panel of Fig. \ref{fig:12}), with perhaps only 
the P$_n$ = 0.0 model mildly discounted, rendering problematic for this EGP 
the constraint on the degree of heat redistribution from this one point alone.  
As Fig. \ref{fig:6} suggests, values of P$_n$ of 0.0, 0.1, 0.2, and 0.3
do not fit the TrES-1 data point at 4.5 \mic to at least 2-$\sigma$, while values of P$_n$
of 0.0 and 0.1 do not fit the TrES-1 data point at 8.0 \mic to the 2-$\sigma$ level.  
The other values of P$_n$ can not be excluded.  Hence, we conclude that
while some heat redistribution by winds to the nightside is definitely indicated 
for TrES-1 and HD209458b, the degree of redistribution is harder to constrain,
with a slight bias towards the larger values of P$_n$.  

The steep slope from 4.5 \mic to 8.0 \mic is best explained by the rise to a peak near $\sim$10 \mic,
which in our models is a natural consequence of the relative strength and positions
of water bands longward of $\sim$5.5 \mic (Fig. \ref{fig:11}).  Note that
without a large water abundance, none of the data nor  
their ratios would make collective sense.  This was the conclusion of 
Burrows, Hubeny, \& Sudarsky (2005), which we reconfirm here.  Furthermore,
the presence of the 4.67 \mic band of CO is indicated by the depth of the 4.5-\mic feature of TrES-1,
but due to the fact that this IRAC-2 band measurement perforce sums over steeply rising fluxes in
regions of the spectrum that bracket the 4.67 \mic feature, and the fact that with
reasonable abundances the band is saturated, almost nothing can be said about 
the CO abundance (Burrows, Hubeny, \& Sudarsky 2005).  We predict 
a rise from IRAC-2 to IRAC-1 for all our models, indicative of the peak we generically see just 
shortward of 4.0 \mic.  We also predict a slight peak around 10 \mic, and a plateau from 
$\sim$14 \mic to 30 \mic. The peak near 10 \mic might be discernible for HD189733b using the full capability of 
{\it Spitzer}/IRS.  The predicted plateau seems suggested by the 
comparison between theory and data for the HD189733b 16-\mic and HD209458b MIPS points,
taken together, but mixing objects (as we have been forced to do with only four data points)
is not very satisfying. 

We have calculated a P$_n$ = 0.5 model for HD209458b with $10\times$solar metallicity 
and, contrary to the conclusion of Fortney et al. (2005), we find that the band contrast
ratios are within $\sim$5\% of those with solar abundances.  This is because,
without clouds, the Bond albedos are very low ($\sles$5\%).  Since changing the
metallicity does not change the total stellar light intercepted by the planet
for a given planet radius, the characteristic atmospheric temperatures are 
similar.  What is more, we find that the $T/P$ profiles are also similar,
with the result that the fluxes and contrast ratios are little altered.  We have not been able
to trace the origin of the difference between our results for higher metallicities
and those of Fortney et al. (2005).  However, we interpret the very weak metallicity dependence 
of the contrast ratios at secondary eclipse for EGP models without clouds  
that we find theoretically to indicate that the metallicity may well be supersolar and large. 
However, by the same token, we conclude that the metallicity can not  
easily be constrained nor measured by secondary eclipse data alone.  Cloud models,
which we expect may be relevant for HD209458b alone among the three EGPs (Fortney et al. 2003;
Sudarsky, Burrows, \& Hubeny 2003), may well change this conclusion
and variations in the C/O ratio, while we do not see any need at this time to invoke
them to fit the four {\it Spitzer} data points, are still of interest (Seager et al. 2005). 

\begin{figure}
\epsscale{1.00}
%\vspace*{-0.7in}
\plotone{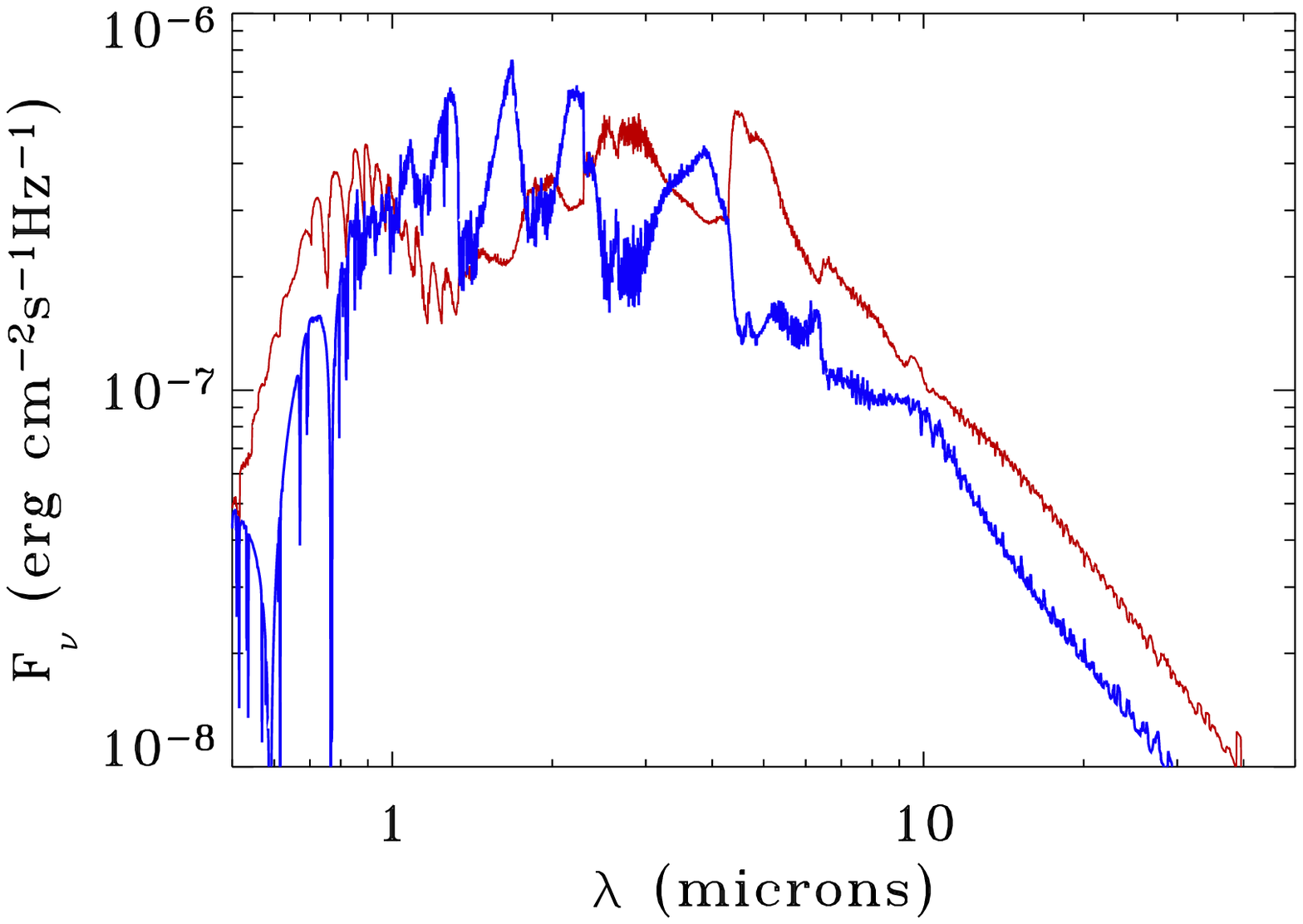}
%\vspace*{-0.2in}
\caption{Two examples of theoretical flux spectra 
(F$_{\nu}$, in erg cm$^{-2}$ s$^{-1}$ Hz$^{-1}$) from the surface of the close-in
EGP OGLE-TR56b from the optical to 30 \mic, with (red line) and without (blue line)
TiO and VO in its upper atmosphere.  For both models, P$_n$ has been 
set equal to 0.0. The differences due to the possible effects of a strong
optical absorber placed in the upper atmosphere of a close-in, hot EGP in the optical, near-IR,
and mid-IR are clearly seen.  In the toy model with ``TiO/VO" (red line), a stratospheric inversion
is produced, the mid-IR and optical are enhanced, and the near-IR is suppressed.  Model numbers
were taken from Hubeny, Burrows, \& Sudarsky (2003).  See text for a discussion. 
\label{fig:14}}
\end{figure}

One way to significantly alter the planet/star contrast ratios is to introduce
at altitude a strong absorber in the optical and near-UV, where the incident stellar
flux can be large.  In this way, the upper atmosphere is heated.  The associated reradiated
optical flux is also greater and the $T/P$ profile manifests a ``stratospheric" inversion
(Hubeny, Burrows, \& Sudarsky 2003). Since the mid-IR fluxes originate 
higher up in the atmosphere than where $\tau_{\rm Rosseland}$$\sim$1,
the associated brightness temperatures from 4 \mic to 30 \mic are also enhanced.  
The increase in the emergent fluxes in the optical and mid-IR leads to a corresponding 
suppression in the near-IR ($\sim$1-4 \mic).   This potential mechanism for altering
our baseline predictions in the {\it Spitzer} bands and for suppressing flux in the near-IR,
particularly in the $Z$, $J$, and $K$ bands, should be borne in mind.  Figure \ref{fig:14},
constructed from a theoretical model found in Hubeny, Burrows, \& Sudarsky (2003), depicts an extreme  
version of this effect for P$_n$ = 0 models of OGLE-TR56b with (red curve) and without (blue curve) 
TiO and VO in its upper atmosphere.  In fact, we expect that TiO and VO are both flushed
out of the upper atmosphere by the coldtrap effect, but suppressing this effect 
allows us to make the general point. Note that the bumps near 10 \mic and 4 \mic seen 
in our fiducial models (Figs. \ref{fig:10} -- \ref{fig:12} and Fig. \ref{fig:13}) 
can be altered, shifted (4 \mic), or muted (10 \mic) by this upper-atmosphere 
absorption effect, so if we fail to see these features as predicted 
interesting stratospheric or upper-atmosphere chemistry might be implied. 
Conversely, the presence of these bumps will put useful limits on such 
upper-atmosphere absorbers.

As mentioned in \S\ref{intro}, we employ the latest measurements of the transit 
radii of each of the three EGPs in determining the planet/star contrast levels.  
However, these radii, being transit radii that probe along the chord of the 
planet in the optical (Burrows, Sudarsky, \& Hubbard 2003), are not strictly  
the appropriate radii to use in determining the planet/star contrast ratios. 
They are close, but the total stellar energy intercepted by the planet 
depends upon wavelength and is different in the near-IR water bands, where 
the transit radius should be slightly larger than in the optical (for HD209458b, 
by $\sim$2-3\%;  Fortney et al. 2003).  Furthermore, planetary emission is from a radius 
that is not corrected for by the ``transit radius effect."  For HD20948b, this
radius difference can be 8-10\% (Burrows, Sudarsky, \& Hubbard 2003), while for TrES-1
and HD189733b, due to the lower atmospheric temperatures and higher gravities
the effect is smaller ($\sles$5\%).  The upshot is that the predicted flux ratios
for HD209458b could be smaller by as much as $\sim$15\%, bringing the 24-\mic
MIPS point better in line with our prediction, but introducing further ambiguities
into predictions at all wavelengths until the radius issue is resolved.  
No one doing theoretical secondary eclipse calculations has yet corrected for 
these subtle radius effects. Moreover, even the measured transit radii retain a residual
ambiguity due to the systematic uncertainty in the stellar radius, which could
easily be 5-10\%.  Knutson et al. (2006) argue that measurements of the star,
constraints of stellar evolution theory (Cody \& Sasselov 2002), and the
detailed fits to the HST/STIS transit measurements together 
yield a transit radius for HD209458b with an error
of only $\sim$2\%.  Perhaps, but our same concerns apply to TrES-1 and HD189733b.
In sum, ambiguities in the appropriate radii to employ in comparing secondary
eclipse data with theory remain and slightly compromise their interpretation.

\begin{figure}
\epsscale{1.00}
%\vspace*{-0.7in}
\plotone{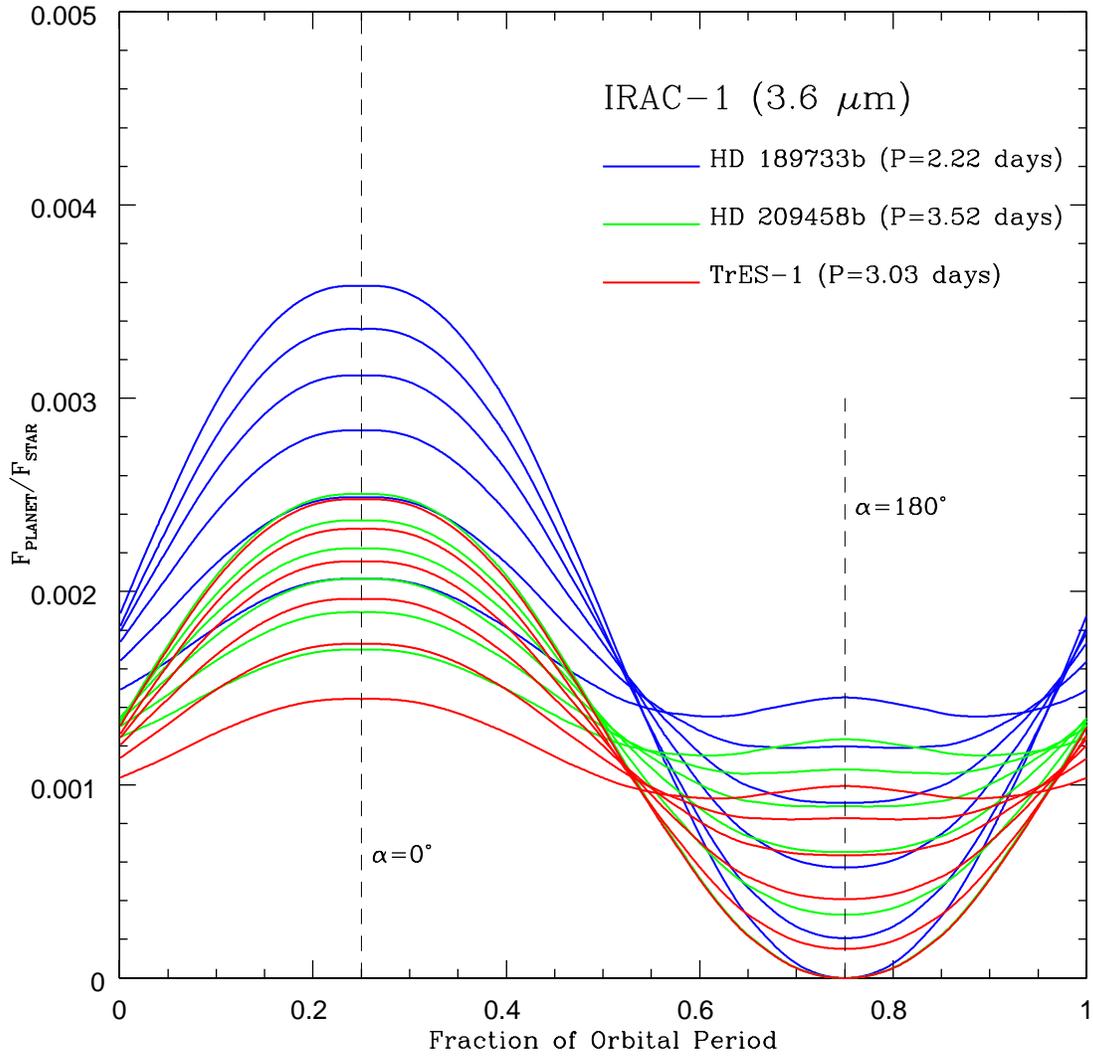}
%\vspace*{-0.2in}
\caption{Light curves versus orbital phase (or orbit fraction) 
of the planet/star flux ratios in the IRAC-1 band centered near 3.6 \mic,
for the transiting EGPs HD189733b (blue), HD209458b (green), and TrES-1 (red)
and for P$_n$s from 0.1 to 0.5, in steps of 0.1.  For a given object or color,
the higher curves are for the lower values of P$_n$ and the dependence upon P$_n$
is monotonic. Indicated by vertical dashed lines are the positions of the $\alpha = 0^{\circ}$ 
(superior conjunction, full phase) and $\alpha = 180^{\circ}$ phases. Given in the legend
are the orbital periods of the three EGPs.  See text for a more detailed discussion. 
\label{fig:5}}
\end{figure}

\begin{figure}
\epsscale{1.00}
%\vspace*{-0.7in}
\plotone{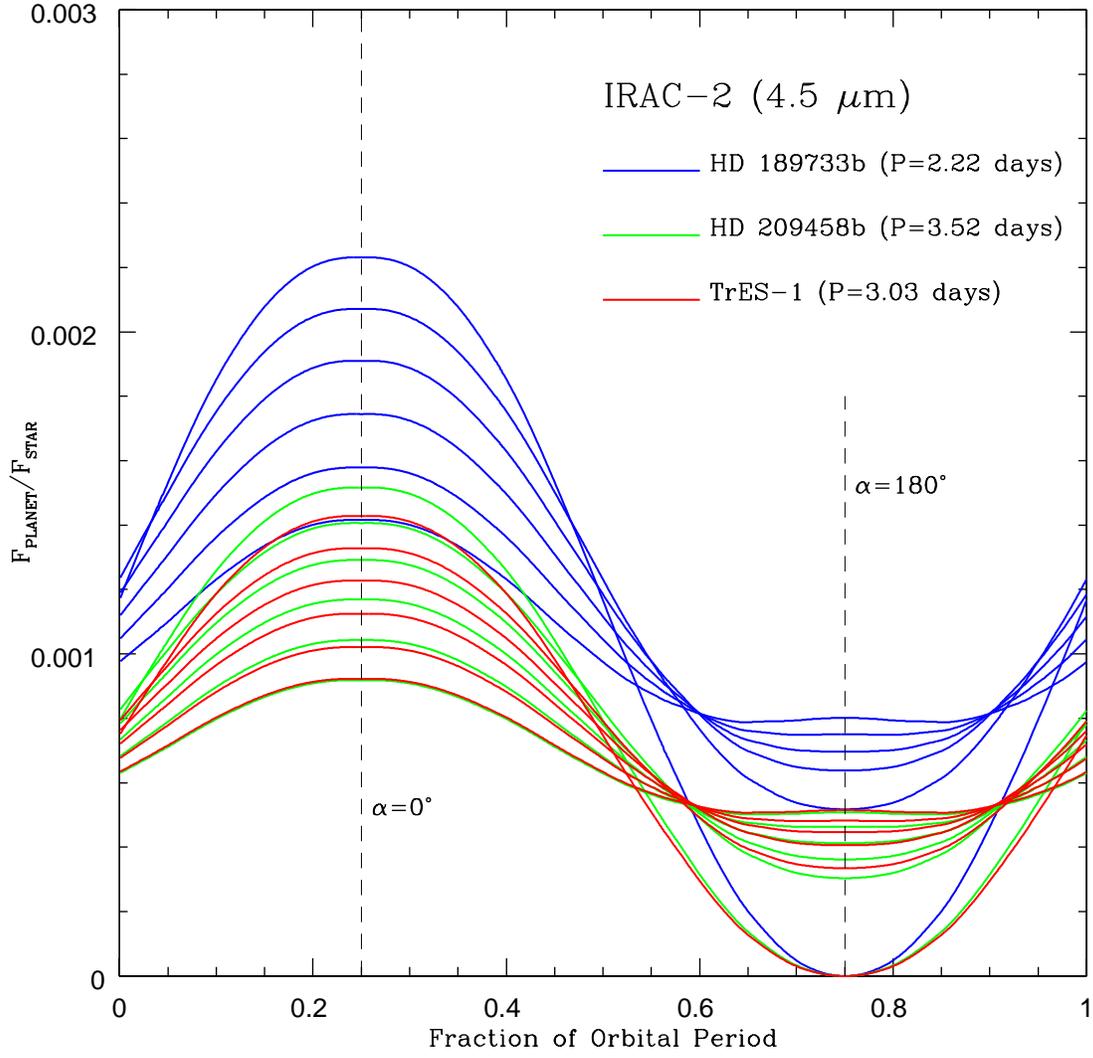}
%\vspace*{-0.2in}
\caption{Same as Fig. \ref{fig:5}, but for the IRAC-2 band centered near
4.5 \mic.
\label{fig:6}}
\end{figure}

\begin{figure}
\epsscale{1.00}
%\vspace*{-0.7in}
\plotone{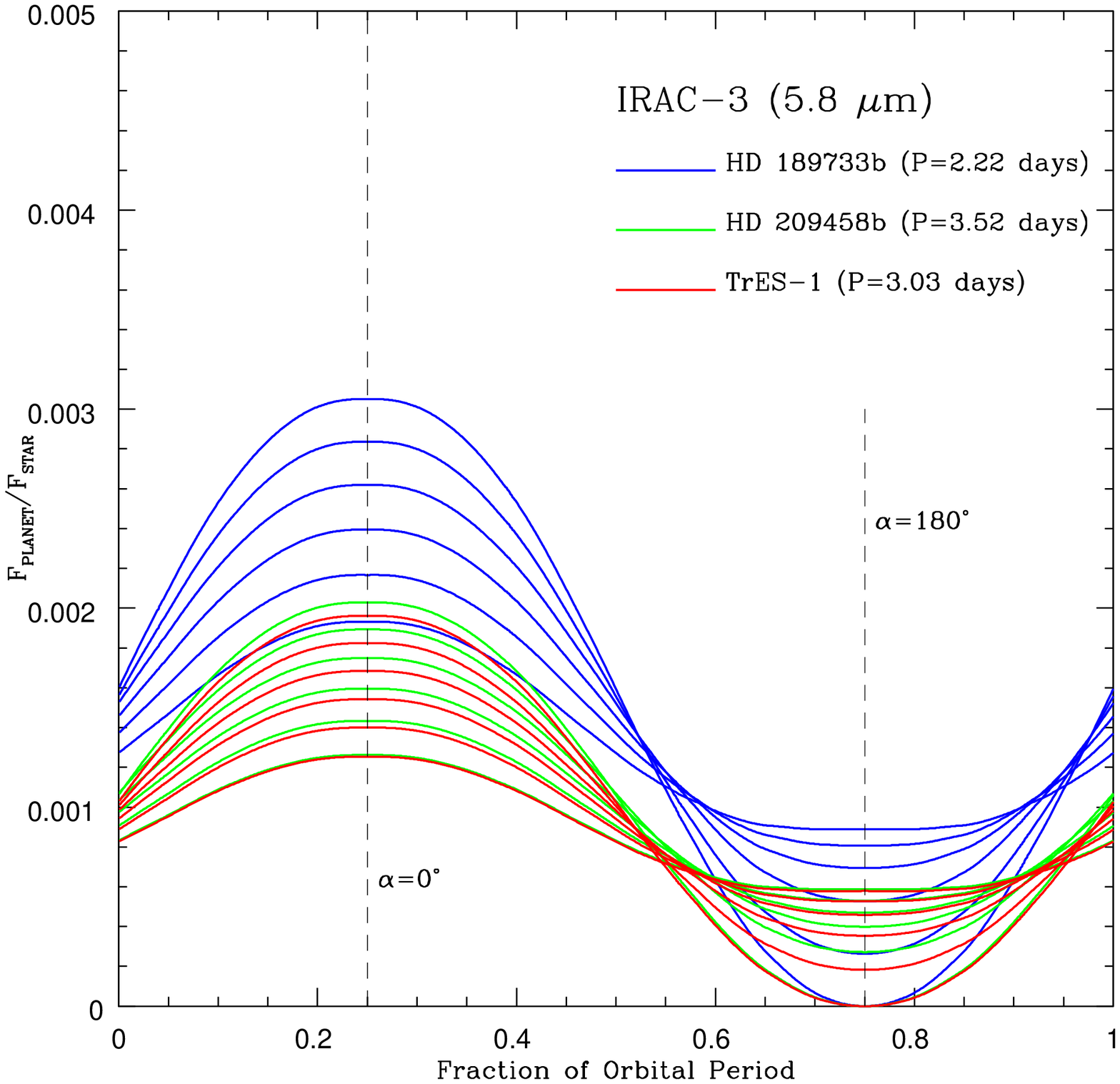}
%\vspace*{-0.2in}
\caption{Same as Fig. \ref{fig:5}, but for the IRAC-3 band centered near
5.8 \mic.  
\label{fig:7}}
\end{figure}

\begin{figure}
\epsscale{1.00}
%\vspace*{-0.7in}
\plotone{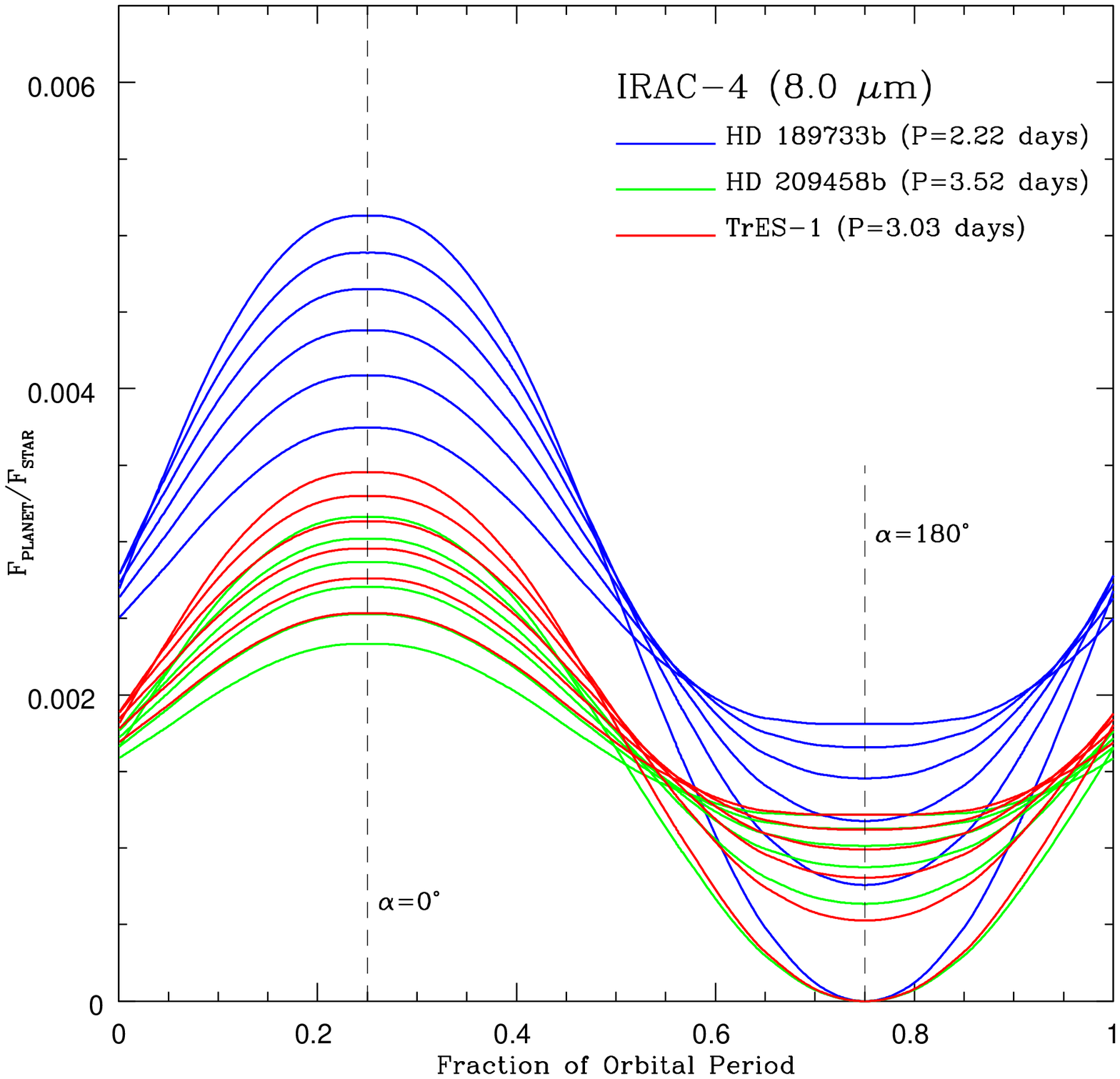}
%\vspace*{-0.2in}
\caption{Same as Fig. \ref{fig:5}, but for the IRAC-4 band centered near
8.0 \mic.
\label{fig:8}}
\end{figure}

\begin{figure}
\epsscale{1.00}
%\vspace*{-0.7in}
\plotone{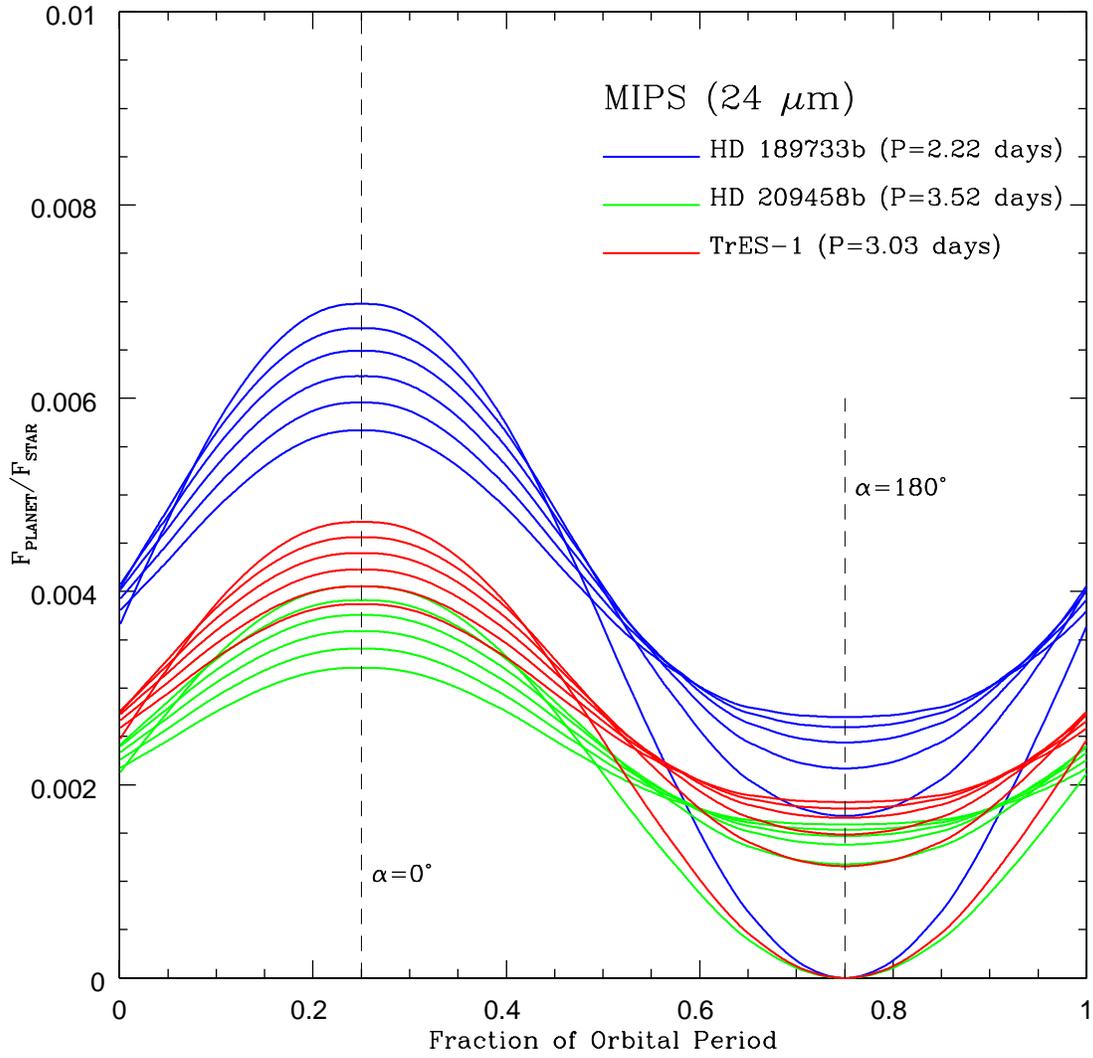}
%\vspace*{-0.2in}
\caption{Same as Fig. \ref{fig:5}, but for the MIPS band centered near
24 \mic.
\label{fig:9}}
\end{figure}

\begin{figure}
\epsscale{1.00}
%\vspace*{-0.7in}
\plotone{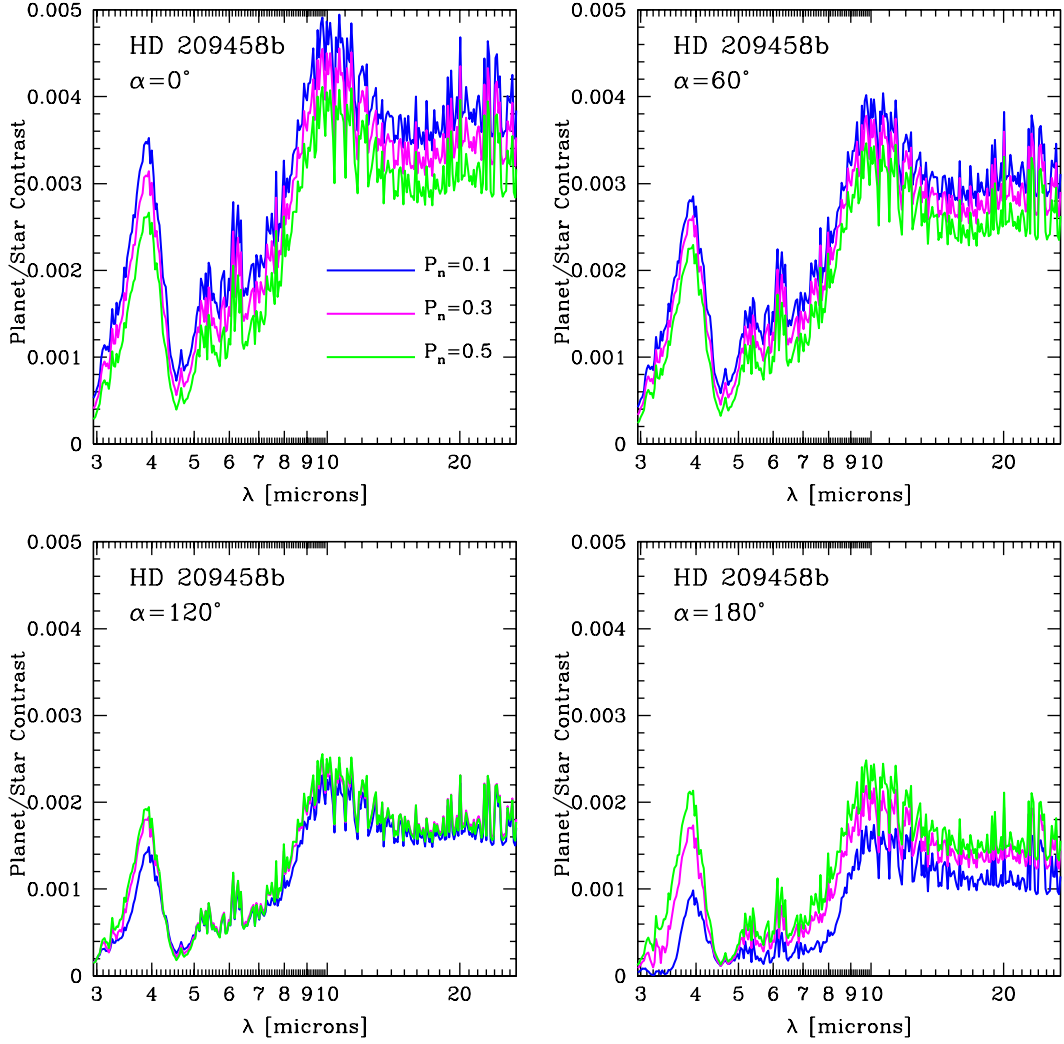}
%\vspace*{-0.2in}
\caption{The HD209458 planet/star flux ratio spectra for 
three representative P$_n$s (0.1, 0.3, 0.5) and four phases 
($\alpha = 0^{\circ}, 60^{\circ}, 120^{\circ}, {\rm and} 180^{\circ}$). 
The wavelength range is 3 \mic to 27 \mic. See text for a 
short discussion of the various features and systematics and 
Figs. \ref{fig:5} through \ref{fig:9} for related theoretical 
results focussed on the IRAC and MIPS bands.
\label{fig:10}}
\end{figure}

\begin{figure}
\epsscale{1.00}
%\vspace*{-0.7in}
\plotone{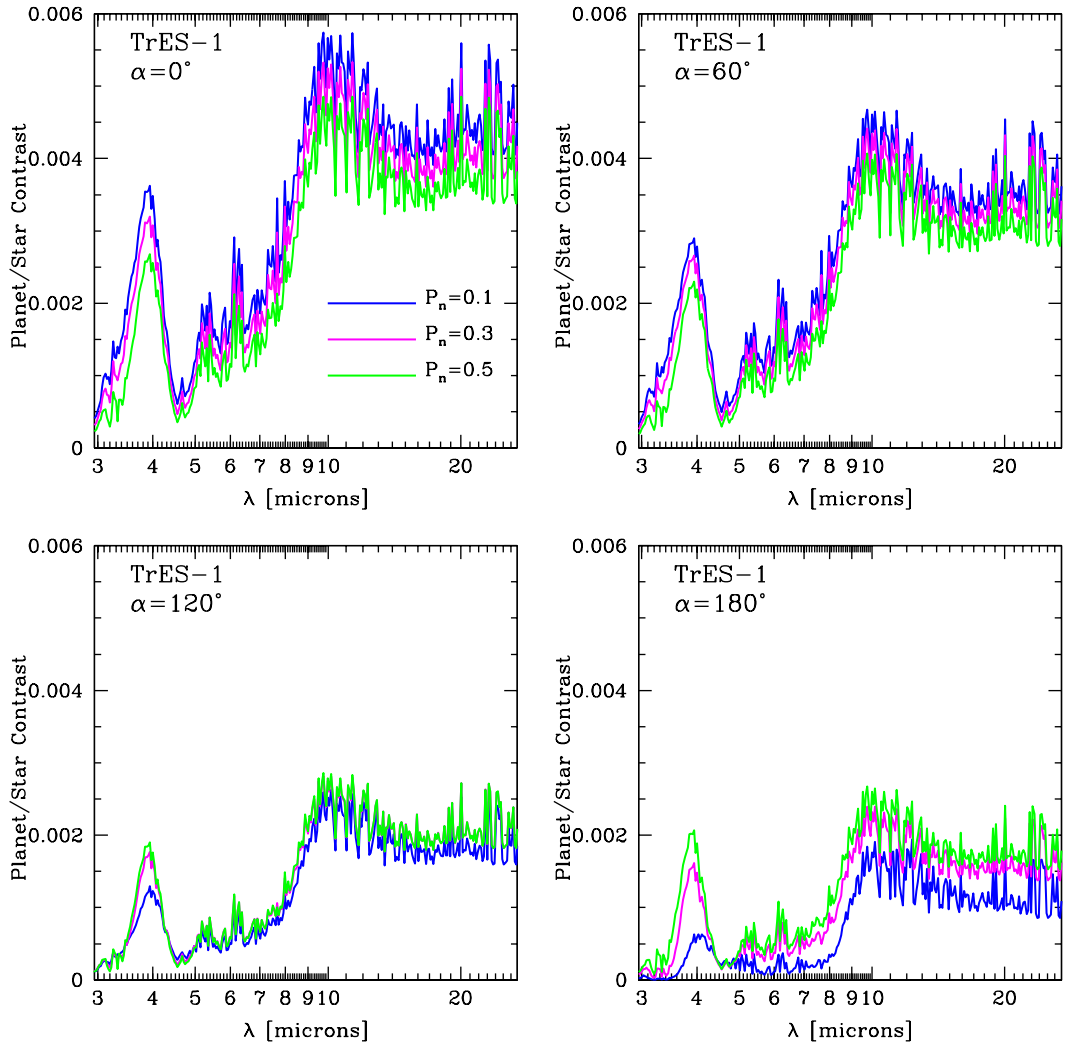}
%\vspace*{-0.2in}
\caption{Same as Fig. \ref{fig:10}, but for TrES-1 and its primary.
\label{fig:11}}
\end{figure}

\begin{figure}
\epsscale{1.00}
%\vspace*{-0.7in}
\plotone{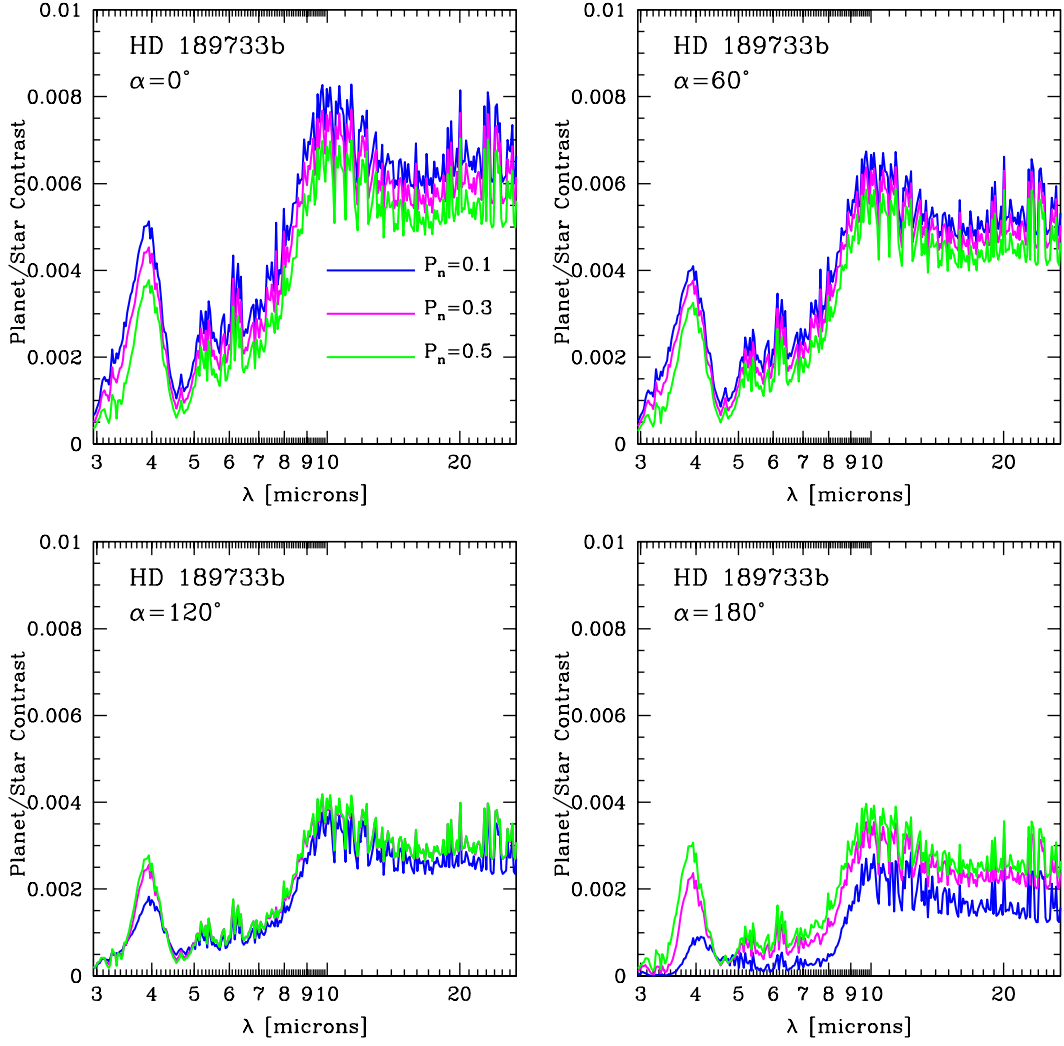}
%\vspace*{-0.2in}
\caption{Same as Fig. \ref{fig:10}, but for the HD189733 EGP/star system.
\label{fig:12}}
\end{figure}

However, that the data and theory we have developed here correspond as well as they
do is gratifying.  In fact, the theory does a good job fitting the
four secondary eclipse data points (Fig. \ref{fig:13}), whatever the ambiguities. 
In addition, there is evidence for redistribtuion 
to the nightside, particularly for TrES-1 and HD209458b,
(though its specific magnitude remains to be determined), and the presence of
H$_2$O and of CO is strongly indicated.  Moreover, we find that the metallicity
dependence of cloud-free models is quite mild, but that ambiguities in the radii remain
to slightly compromise the interpretation of the data.  Due to the greater sensitivity
in the IRAC bands to variations in P$_n$ (Figs. \ref{fig:5} - \ref{fig:8}), such data have greater
potential to determine the degree(s) of redistribution.  Data off secondary eclipse at
other phase angles, and particularly in IRAC-4 (Figs. \ref{fig:10} - \ref{fig:12}),
would further constrain the models, but the most propitious phase angles
in this regard are larger than 90$^{\circ}$ and, hence, will prove very difficult 
to measure.  However, JWST, with its two to three orders-of-magnitude greater
sensitivity in the mid-IR, will be able to measure a large fraction of the 
planetary light curves.  In the shorter term, data, or even upper limits, at the 14 other 
anticipated {\it Spitzer} band points are anxiously awaited.

\acknowledgments

We thank Christopher Sharp, Bill Hubbard, Jaymie Matthews, 
Jonathan Fortney, Mike Cushing, and Drew Milsom for helpful discussions.
This study was supported in part by NASA grant NNG04GL22G and 
through the NASA Astrobiology Institute under Cooperative 
Agreement No. CAN-02-OSS-02 issued through the Office of Space
Science.  The models presented in this paper are available in 
electronic form from the first author upon request.  A web-based 
calculator for determining the optical and near-IR light 
curves of EGPs in wide orbits as a function of orbital phase, 
wavelength, semi-major axis, orbital inclination, and eccentricity
is now available at http://zenith.as.arizona.edu/\~{}burrows/phase/lightcurve.php.
A similar calculator for close-in EGPs is under development.

{}

\clearpage

\begin{deluxetable}{llllllllll}
%\tablewidth{18.2cm}
\tablewidth{17.0cm}
\tablenum{1}
\tabletypesize{\scriptsize}
\tablecaption{Reference Data for Transiting EGPs\label{EGP_tab}}
\tablehead{
\colhead{EGP} & \colhead{Star} & \colhead{M$_{\ast}$ (M$_{\odot}$)} & \colhead{R$_{\ast}$ (R$_{\odot}$)}
& \colhead{$a$ (AU)}
& \colhead{P (days)} & \colhead{M$_{p}$ (\mj)} & \colhead{R$_p$ (R$_{\rm J}$)} & \colhead{Log$_{10}$ $g_p$ (cgs)} & \colhead{[Fe/H]}}
\startdata

HD209458b & G0V  & 1.05 & 1.1 & 0.045 & 3.524 & 0.69 & 1.32 & 2.99 & 0.01\\
HD189733b & K2V  & 0.82 & 0.75 & 0.031 & 2.22 & 1.15 & 1.154 & 3.25 & -0.03 \\
TrES-1 & K0V  & 0.87 & 0.86 & 0.0393 & 3.03 & 0.76 & 1.08 & 3.21 & 0.0 \\

\enddata
\tablerefs{\rj (Jupiter's radius) = $7.149\times 10^4$ km; see J. Schneider's 
Extrasolar Planet Encyclopaedia at http://www.obspm.fr/encycl/encycl.html
and the Carnegie/California compilation at http://exoplanets.org}

\end{deluxetable}

\clearpage

% figure 1

\clearpage

% figure 2

\clearpage

% figure 3

\clearpage

% figure 4

\clearpage

% figure 5

\clearpage

% figure 6

\clearpage

% figure 7

\clearpage

% figure 8

\clearpage

% figure 9

\clearpage

% figure 10

\clearpage

% figure 11

\clearpage

% figure 12

\clearpage

% figure 13

\clearpage

% figure 14


\begin{thebibliography}{}

\bibitem[Allard et al. 2003]{allard} Allard, F., Baraffe, I.,
Chabrier, G., Barman, T.S., \& Hauschildt, P.H. 2003, 
in ``Scientific Frontiers in Research 
on Extrasolar Planets," ASP Conf. Series vol. 28 (PASP v.294), 
eds. D. Deming \& S. Seager, p. 483

%\bibitem[Alonso \etal 2004]{alonso} Alonso, R. \etal 2004, \apj, 613, L153 % TrES-1

%\bibitem[Antonello \& Ruiz]{AntonelloRuiz02} Antonello, E. \&
%Ruiz, S. M. 2002, {\it The Corot Mission},
%\verb"http://www.astrsp-mrs.fr/projects/corot/corotmission.ps"

\bibitem[Bakos et al. 2006]{bakos} Bakos, G. \'A. et al. 2006, 
submitted to \apj, (astro-ph/0603291) %refined parameters for HD189733b: 1.154 Rj

%\bibitem[Baraffe et al. 2003]{baraffe2} Baraffe, I., 
%Chabrier, G., Barman, T.S., Allard, F., \& Hauschildt, P.H.
%2003, \aap, 402, 701

%\bibitem[Baraffe et al. 2004]{baraffe4} Baraffe, I., Selsis, F., Chabrier, G., 
%Barman, T.S., Allard, F., \& Hauschildt, P.H., \& Lammer, H. 2004, astro-ph/0404101 % evaporation

\bibitem[Barman, Hauschildt, \& Allard 2005]{barman} Barman, T.S., Hauschildt, P.H., \& Allard, F. 
2005, \apj, 632, 1132

%\bibitem[Bond et al. 2004]{bond} Bond, I.A., et al. 2004, astro-ph/0404309 % OGLE Microlensing

%\bibitem[Bodenheimer, Lin, \& Mardling 2001]{boden} Bodenheimer, P.,
%Lin, D.N.C., Mardling, R.A. 2001, \apj, 548, 466

%\bibitem[Bodenheimer, Laughlin, \& Lin 2003]{boden2} Bodenheimer, P.,
%Laughlin, G., \& Lin, D.N.C. 2003, \apj, 592, 555

%\bibitem[Bouchy et al. 2004]{bouchy} Bouchy, F., Pont, F., Santos, F.C., Melo, C., 
%Mayor, M., Queloz, D., \& Udry, S. 2004, astro-ph/0404264 % OGLE-TR-113 and OGLE-TR-132

%\bibitem[Brown et al. 2001]{brown01} Brown, T. M., Charbonneau, D.,
%Gilliland, R.L., Noyes, R.W., \& Burrows, A. 2001, \apj, 552, 699

\bibitem[Burkert et al. (2005)]{burkert} Burkert, A., Lin, D.N.C., 
Bodenheimer, P., Jones, C., \& Yorke, H. 2005, \apj, 618, 512

%\bibitem[Burrows \& Lunine 1995]{burlun} Burrows, A. \& Lunine, J.I. 1995, Nature, 378, 333

%\bibitem[Burrows \etal 1997]{Burrows97} Burrows, A., Marley, M.,
%Hubbard, W. B., Lunine, J. I., Guillot, T., Saumon, D., Freedman, R.,
%Sudarsky, D., \& Sharp, C. 1997, \apj, 491, 856

%\bibitem[Burrows \& Sharp 1999]{BurrowsSharp99} Burrows, A. \&
%Sharp, C. M. 1999, \apj, 512, 843

\bibitem[Burrows \etal 2000]{Burrows00} Burrows, A., Guillot, T.,
Hubbard, W. B., Marley, M. S., Saumon, D., Lunine, J. I.,
\& Sudarsky, D. 2000, \apj, 534, 97 

%\bibitem[Burrows \etal 2001]{bur01} Burrows, A., Hubbard, W.B., 
%Lunine, J.I., and Liebert, J. 2001, Rev. Mod. Phys., 73, 719

\bibitem[Burrows, Sudarsky, \& Hubbard 2003]{Burrows03} Burrows, A., Sudarsky, \& Hubbard, W.B. 2003, \apj,
594, 545 (BSH)

\bibitem[Burrows,~Sudarsky,~\&~Hubeny 2003]{holt} Burrows, A., Sudarsky, D.,
\& Hubeny, I. 2003, published in the proceedings of the 14th Annual Astrophysics Conference in Maryland ``The Search
for Other Worlds," eds. S. Holt and D. Deming, (AIP Conference Proceedings),
held in College Park, MD, October 13-14, 2003, p. 143.

\bibitem[Burrows,~Sudarsky,~\&~Hubeny 2004]{bur04a} Burrows, A., Sudarsky, D., \& Hubeny, I. 2004,
\apj, 609, 407 % Wide-separation

\bibitem[Burrows et al. 2004]{bur2004f} Burrows, A., Hubeny, I., Hubbard, W.B.,
Sudarsky, D., \& Fortney, J.J. 2004 \apj, 610, L53

\bibitem[Burrows 2005]{bur05} Burrows, A. 2005, Nature, 433, 261

\bibitem[Burrows et al. 2005]{bur2005} Burrows, A., Hubeny, I., \& Sudarsky, D.,
2005 \apj, 625, L135

%\bibitem[Chabrier et al. 2004]{chab2004} Chabrier, G., Barman, T., 
%Baraffe, I., Allard, F., \& Hauschildt, P.H. 2004, \apj, 603, L53

\bibitem[Charbonneau \etal 2000]{Charbonneau00} Charbonneau, D., 
Brown, T. M., Latham, D. W., \& Mayor, M. 2000, \apjl, 529, L45

\bibitem[Charbonneau \etal 2002]{Charbonneau02} Charbonneau, D.,
Brown, T. M., Noyes, R. W., \& Gilliland, R. L. 2002, \apj, 568, 377

\bibitem[Charbonneau \etal 2005]{char05} Charbonneau, D. \etal 2005, \apj, 626, 523 
%Detection of Thermal Emission from an Extrasolar Planet

\bibitem[Charbonneau et al. 2006]{ppv} Charbonneau, D. Brown, T.M., Burrows, A., \& Laughlin, G. 2006, 
to be published in ``Protostars and Planets V," ed. B. Reipurth and D. Jewitt 
(University of Arizona Press), astro-ph/0603376

\bibitem[Cho \etal 2003]{Cho02} Cho, J. Y-K., Menou, K., Hansen, B. M. S.,
\& Seager, S. 2003, \apj, 587, L117

\bibitem[Cody \& Sasselov 2002]{CodySasselov02} Cody, A. M. \&
Sasselov, D. D. 2002, \apj, 569, 451

\bibitem[Cooper \& Showman 2005]{cooper} Cooper, C.S. \& Showman, A.P. 2005, 
\apj, 629, L45 (astro-ph/0502476)

\bibitem[Deming \etal 2005]{deming05} Deming, D., Seager, S., Richardson, L.J., \& Harrington, J.,
2005, Nature, 434, 740
% Infrared radiation from an extrasolar planet

%\bibitem[Deming \etal 2005b]{deming05b} Deming, D., Brown, T.M., 
%Charbonneau, D., Harrington, J., \& Richardson, L.J. 2005,
%\apj, 622, 1149 % Search for CO

\bibitem[Deming \etal(2006)]{deming06} Deming, D., Harrington, J.,
Seager, S., Richardson, L.R. 2006, submitted to \apj

%\bibitem[Dyudina et al. 2005]{dyudina} Dyudina, U.A., Sackett, P.D., Bayliss, D.D.R., Seager, 
%S., Porco, C.C., Throop, H.B., \& Dones, L. 2005, \apj, 618, 973

\bibitem[Fortney et al. 2003]{fort03} Fortney, J.J., Sudarsky, D., Hubeny, I., Cooper, C.S.,
Hubbard, W.B., Burrows, A., \& Lunine, J.I. 2003, \apj, 589, 615

\bibitem[Fortney et al. 2005]{fort2005} Fortney, J.J., Marley, M.S., Lodders, K., 
Saumon, D., Freedman, R.S. 2005, \apj, 627, L69 % Comparative TrES-1 and HD209458b

%\bibitem[Fortney et al. 2005b]{fort2005b} Fortney, J.J., Saumon, D., Marley, M.S., Lodders, K., 
%\& Freedman, R.S. 2005, accepted to \apj, astro-ph/0507422 % HD149026b and HD189733b

%\bibitem[Guillot \etal 1996]{Guillot96} Guillot, T., Burrows, A.,
%Hubbard, W. B., Lunine, J. I., \& Saumon, D. 1996, \apj, 459, 35

\bibitem[Guillot \& Showman 2002]{GuillotShowman02} Guillot, T. \&
Showman, A. P. 2002, \aap, 385, 156

%\bibitem[Henry \etal 2000]{Henry00} Henry, G., Marcy, G. W., Butler, R. P.,
%\& Vogt, S. S. 2000, \apjl, 529, L41

%\bibitem[Hubbard \etal 2001]{Hubbard01} Hubbard, W. B., Fortney, J. F.,
%Lunine, J. I., Burrows, A., Sudarsky, D., \& Pinto, P. A. 2001,
%\apj, 560, 413

%\bibitem[Hubeny 1988]{Hubeny88} Hubeny, I. 1988, Computer Physics Comm., 52, 103

%\bibitem[Hubeny \& Lanz 1995]{HubenyLanz95} Hubeny, I. \& Lanz, T. 1995,
%\apj, 439, 875

\bibitem[Hubeny, Burrows, \& Sudarsky 2003]{Hubeny03} Hubeny, I., Burrows, A., \& Sudarsky, D. 2003,
\apj, 594, 1011

\bibitem[Iro, B\'ezard, \& Guillot 2005]{niro} Iro, N., B\'ezard, B., 
\& Guillot, T. 2005, \aap, 436, 719

\bibitem[Knutson et al. 2006]{knutson} Knutson, H., Charbonneau, D., Noyes, R.W., Brown, T.M.,
Gilliland, R.L. 2006, submitted to \apj, astro-ph/0603542

%\bibitem[Koch \etal 1998]{Koch98} Koch, D., Borucki, W., Webster, L.,
%Dunham, E., Jenkins, J., Marrion, J., \& Reitsema, H. 1998, SPIE
%Conference 3356:
%{\it Space Telescopes and Instruments V}, 599

%\bibitem[Konacki et al. 2003a]{konacki2003a} Konacki, M., Torres, G., Jha, S., \& Sasselov, D. 2003,
%Nature, 421, 507 % OGLE-TR-56b

%\bibitem[Konacki et al. 2003b]{konacki2003b} Konacki, M., Torres, G., Sasselov, D., \& Jha, S. 2003,
%\apj, 597, 1076 % OGLE-TR-10

%\bibitem[Konacki et al. 2004]{konacki04} Konacki, M., et al. 2004, \apj, 609, L37 (astro-ph/0404541) % OGLE-TR-113

\bibitem[Kurucz 1994]{Kurucz94} Kurucz, R. 1994, {\it Kurucz CD-ROM
No. 19}, (Cambridge: Smithsonian Astrophysical Observatory)

%\bibitem[Laughlin et al. 2005]{laugh} Laughlin, G., Wolf, A., Vanmunster, T., Bodenheimer, P.,
%Fischer, D., Marcy, G., Butler, P., \& Vogt, S. 2005, preprint % TrES-1 data

%\bibitem[Lecavalier~des~Etangs et al. 2004]{lecav} Lecavelier des Etangs, A. et al. 2004, \aap, 418, L1

%\bibitem[Marcy \& Butler 1996]{mb96} Marcy, G.W. \& Butler, R.P. 1996, \apj, 464, L147

%\bibitem[Marcy \& Butler 1998]{mb98} Marcy, G.W. \& Butler, R.P. 1998, \araa, 36, 57

%\bibitem[Marcy, Cochran, \& Mayor 2000]{mcm} Marcy, G.W., W. Cochran, W., \& Mayor, M. 2000,
%in {\it Protostars and Planets IV},
%ed. V. Mannings, A.P. Boss, and S.S. Russell (Tucson: The University of Arizona Press), p. 1285-1311

%\bibitem[Mayor \& Queloz 1995]{mayor} Mayor, M. \& Queloz, D. 1995, Nature, 378, 355

%\bibitem[Mazeh \etal 2000]{Mazeh00} Mazeh, T., Naef, D., Torres, G.,
%\etal 2000, \apjl, 532, L55

\bibitem[Menou \etal 2002]{Menou02} Menou, K, Cho, J. Y-K., Hansen, B. M. S.,
\& Seager, S. 2003, \apj, 587, L113. 

%\bibitem[Perryman 1997]{perry} Perryman, M.A.C. 1997, The {\it Hipparcos} and 
%{\it Tycho} Catalogues (ESA SP-1200, Noordwijk:ESA)

%\bibitem[Pont et al. 2004]{pont2004} Pont, F., Bouchy, F., Queloz, D., Santos, N. C., Melo, C., Mayor, M., \& Udry, S.
%2004, \aap, 426, L15 % OGLE-TR-111b

\bibitem[Richardson, Deming, \& Seager 2003]{richard} Richardson, L.J., Deming, D., 
\& Seager, S. 2003, \apj, 597, 581 % HD209458b: Strong Limits on 2.2 micron

\bibitem[Rowe et al. 2006]{rowe} Rowe, J.F., Matthews, J.M., Seager, S., Kuschnig, R., Guenther, D.B.,
Moffat, A.F.J., Rucinski, S.M., Sasselov, D., Walker, G.A.H., \& Weiss, W.W. 2006,
accepted to \apj, 645 (astro-ph/0603410)% A_B < 0.375, A_g < 0.25

%\bibitem[Saumon, Chabrier, \& Van Horn 1995]{sc95} Saumon, D., Chabrier, G., \& Van Horn, H. 1995,  \apjs, 99, 713

%\bibitem[Sasselov (2003)]{sasselov03} Sasselov, D. 2003, \apj, 596, 1327 % OGLE-TR56b

\bibitem[Seager et al. 2005]{seager2005} Seager, S., Richardson, L.J., 
Hansen, B.M.S., Menou, K., Cho, J.Y.-K., \& Deming, D. 2005, \apj, 632, 1122 
% "On the Dayside Thermal Emission of Hot Jupiters"

\bibitem[Showman \& Guillot 2002]{ShowmanGuillot02} Showman, A. P.
\& Guillot, T. 2002, \aap, 385, 166

%\bibitem[Smith \& Hunten 1990]{smith90} Smith, G. R. \& Hunten, D. M. 1990,
%Rev. Geophys., 28, 117

%\bibitem[Sozzetti et al. 2004]{sozzetti} Sozzetti, A. et al. 2004, \apj, 616, L167 (astro-ph/0410483) % High-res. TrES-1

\bibitem[Sudarsky \etal 2000]{Sudarsky00} Sudarsky, D., Burrows, A.,
\& Pinto, P. 2000, \apj, 538, 885

\bibitem[Sudarsky, Burrows, \& Hubeny 2003]{sud02}
Sudarsky, D., Burrows, A., \& Hubeny, I. 2003, \apj, 588, 1121

\bibitem[Sudarsky, Burrows, Hubeny, \& Li 2005]{sud05}
Sudarsky, D., Burrows, A., Hubeny, I., \& Li, A. 2005, \apj, 627, 520 
(see http://zenith.as.arizona.edu/\~{}burrows/phase/lightcurve.php) 

%\bibitem[Torres et al. 2005]{torres} Torres, G., Konacki, M., Sasselov, 
%D., \& Jha, S. 2005,  \apj, 619, 558 (astro-ph/0310114) % OGLE-TR56

\bibitem[Trauger et al. 2000]{Trauger00} Trauger, J. \etal 2000,
BAAS, 197, \# 49.07, 1486

\bibitem[Trauger, Hull, and Redding 2001]{Trauger01} Trauger, J., Hull, A.B.,
\& Redding, D.A. 2001, BAAS, 199, \# 86.04, 1431


\bibitem[Vidal-Madjar et al. 2003]{vidal} Vidal-Madjar, A., Lecavelier des Etangs, A., D\'esert, J.-M.,
Ballester, G.E., Ferlet, R., H\'ebrand, G., \& Mayor, M. 2003, Nature, 422, 143

%\bibitem[Vidal-Madjar et al. 2004]{vidal2} Vidal-Madjar, A., D\'esert, J.-M., 
%Lecavelier des Etangs, A., H\'ebrard, G., Ballester, G.E., Ehrenreich, D., 
%Ferlet, R., McConnell, J.C., Mayor, M., Parkinson, C. D. 2004, \apj, 604, L69

\bibitem[Williams et al. 2006]{will 2006} Williams, P.K.G., Charbonneau, D., Cooper, 
C.S., Showman, A.P., Fortney, J.J., submitted to \apj, astro-ph/0601092 % "Resolving the Surfaces..."

\end{thebibliography}
\end{document}